\documentclass[aps,pra,reprint]{revtex4-1}
\usepackage{graphics}
\usepackage{epsfig}
\usepackage{amsmath}
\usepackage{amssymb}
\DeclareMathOperator{\rk}{rk}
\DeclareMathOperator{\Tr}{Tr}
\DeclareMathOperator{\Spect}{Spect}
\newcommand{\td}{\tilde{d}}
\newcommand{\tg}{\tilde{g}}
\newcommand{\tw}{\tilde{w}}
\newcommand{\cket}[1]{\vert #1 \rangle}
\newcommand{\bra}[1]{\langle #1 \vert}
\newcommand{\bracket}[1]{\langle #1 \rangle}
\newcommand{\Id}{\mathbb{I}}

\begin{document}
\title{A study of separability criteria for mixed three-qubit states}
\author{Szil\'ard Szalay}
\email{szalay@phy.bme.hu}
\affiliation{
Department of Theoretical Physics, 
Institute of Physics, 
Budapest University of Technology and Economics, 
H-1111 Budapest, Budafoki \'ut 8, Hungary}
\date{\today}
\begin{abstract}
We study the noisy GHZ-W mixture.
We demonstrate some necessary but not sufficient criteria
for different classes of separability of these states.
It turns out that
the partial transposition criterion of Peres \cite{PeresCrit}
and the criteria of G\"uhne and Seevinck \cite{GuhneSevinckCrit} dealing with matrix elements
are the strongest ones for different separability classes of this 2 parameter state.
As a new result we determine a set of entangled states of positive partial transpose.
\end{abstract}

\maketitle{}

\section{Introduction}

The notion of entanglement \cite{Horodecki4} is the characteristic trait of quantum mechanics.
It serves as a resource for Quantum Information Theory \cite{NielssenChuang},
a relatively new field of science
dealing with the properties, characterisation and applications
(e.g.~quantum computation \cite{NielssenChuang}, quantum teleportation \cite{Teleport})
of the nonlocal behavior of entangled quantum states.
A variety of methods of Quantum Information Theory uses pure entangled states of a quantum system
which can be easily prepared and which are easy to use to get nonclassical results.
However, in a laboratory one can not get rid of the interaction with the environment perfectly,
thus the separable compound state of the system and the environment evolves into an entangled one,
the prepared pure state of the system evolves into a mixed one.

Generally it is a difficult question to decide whether a mixed state is entangled or not.
\cite{Horodecki4,GuhneTothEntDet}
A density operator representing the state
acting on an
$\mathcal{H}=\mathcal{H}^A\otimes\mathcal{H}^B$
composite Hilbert space by definition is \emph{separable} \cite{WernerSep}
when it can be written as a convex combination of products of local density operators,
i.e.~if there exists a decomposition of the form
\begin{equation} 
\label{sep} 
\varrho=\sum_i p_i\varrho^A_i\otimes\varrho^B_i, 
\end{equation}
where $0\leq p_i$, $\sum_ip_i=1$,
and $\varrho^A_i$ and $\varrho^B_i$ are
positive operators of trace one
acting on subsystems $\mathcal{H}^A$ and $\mathcal{H}^B$ respectively.
Classical correlations can give rise only to separable states in the sense of Eq.~(\ref{sep}). \cite{WernerSep}

A decomposition like the one of Eq.~(\ref{sep}) is not unique, 
and it is difficult to decide whether for a given density operator such a decomposition exists at all.
One can make some observations for separable pure states
which can be extended to mixed states with the help of convex calculus.
The \emph{separability criteria} obtained in this way are \emph{necessary but not sufficient} ones.
(Or equivalently sufficient but not necessary \emph{criteria of entanglement.})
On the other hand one can construct \emph{necessary and sufficient} criteria 
using sophisticated mathematical methods.~\cite{PeresHorodeckiCrit}
Unfortunately these criteria are difficult to use for general density matrices
and only the necessary but not sufficient criteria can be used in practice.~\cite{BengtssonZyczkowski}
In this paper we calculate explicitly some of the necessary but not sufficient criteria of separability
for a particular two parameter mixture of \emph{three-qubit} density matrices.
The form of these density matrices are simple enough
to calculate explicitly the set of states for which these criteria hold.

The organization of this paper is as follows.
First of all in Section \ref{secClasses} we briefly review 
the \emph{separability classes of three-qubit mixed states}
using the notions $\alpha_k$-separability and $k$-separability.
In Section \ref{secRho} we introduce our parametrized permutation-invariant family of \emph{three-qubit} density matrices
and make some observations about the separability class structure of \emph{permutation-invariant} three-qubit mixed states.
After having set the stage, in the next sections we investigate some criteria for separability classes.
First in Section \ref{secTwopart} we consider our quantum-state as a $2\times4$ qubit-qudit system
and we recall and use some \emph{bipartite} separability criteria,
namely the \emph{majorization} and the \emph{entropy criteria} related to the notion of mixedness of the subsystems
(Sections \ref{secMaj} and \ref{secEntr} respectively),
the \emph{partial transposition} and the \emph{reduction criteria} which are particular cases of the positive map criterion
(Sections \ref{secPPT} and \ref{secRed} respectively),
and the \emph{reshuffling criterion} which 
in addition to the partial transposition criterion
is the other one of the two independent permutation criteria for two-partite systems 
(Section \ref{secResh}).
As a next step in Section \ref{secThreepart} we consider our quantum-state as a proper $2\times2\times2$ three-qubit system
and investigate some \emph{three-partite} criteria for separability classes.
In Section \ref{secPerm} we recall the \emph{permutation criteria} for permutation-invariant three-qubit case
giving rise to \emph{another reshuffling criterion}.
Then we use some criteria using the \emph{expectation value of local spin-observables} (Section \ref{secSpin}),
\emph{swap operators} (Section \ref{secHub})
and \emph{explicit expressions of matrix elements} (Section \ref{secMatrix}).
The latter makes it possible to determine a set of entangled states of positive partial transpose.
In Section \ref{secClass1} we investigate the \emph{SLOCC classes of fully entangled states}.
A \emph{summary} is given in Section \ref{secConcl}.
The \emph{explicit form} of the corresponding matrices,
some examples for permutation-invariant states of special separability classes
and a detailed calculation of \emph{Wootters-concurrence} of the corresponding two-qubit subsystems are left to 
Appendices \ref{secMatrices}, \ref{secExampl} and \ref{secWoott} respectively.

\section{Separability classes}
\label{secClasses}

A two-partite mixed state can be either separable or entangled, 
depending on the existence of a decomposition as given by Eq.~(\ref{sep}).
However, the structure of separability classes can be very complex even for three subsystems.
To get the adequate generalization of Eq.~(\ref{sep}) we recall the definitions of $k$-separability and $\alpha_k$-separability
as given in Ref.~\cite{SeevinckUffink}.

Consider an $N$-qubit system with Hilbert space $\mathcal{H}=\mathcal{H}^1\otimes\mathcal{H}^2\otimes\dots\otimes\mathcal{H}^N$
and denote the full set of states for this system as $\mathcal{D}_N$.
Let $\alpha_k=(S_1,S_2,\dots,S_k)$ denote a \emph{partition} of the labels $\{1,2,\dots,N\}$ into $k\leq N$ disjoint nonempty subsets $S_n$.
A density matrix is \emph{$\alpha_k$-separable}, i.e.~separable under the \emph{particular} $k$-partite split $\alpha_k$,
if and only if it can be written as a convex combination of product states
with respect to the split $\alpha_k$. 
We denote the set of these states as $\mathcal{D}_N^{\alpha_k}$:
\begin{equation}
\label{sepak}
\varrho\in\mathcal{D}_N^{\alpha_k}:\qquad
\varrho=\sum_ip_i\otimes_{n=1}^k\varrho_i^{S_n},
\end{equation}
where $0\leq p_i$, $\sum_ip_i=1$ and
$\varrho_i^{S_n}$ is a density operator
 of the subsystem corresponding to $S_n$ in the split $\alpha_k$
(i.e.~acting on $\mathcal{H}^{S_n}=\otimes_{a\in S_n}\mathcal{H}^a$).
More generally, \emph{for a given} $k$ we can consider states which can be written as a mixture of
$\alpha_k^{(i)}$-separable states for generally different $\alpha_k^{(i)}$ splits.
These states are called \emph{$k$-separable states} and denoted as $\mathcal{D}_N^{k-\text{sep}}$:
\begin{equation}
\label{sepk}
\varrho\in\mathcal{D}_N^{k-\text{sep}}:\qquad
\varrho=\sum_ip_i\otimes_{n=1}^k\varrho_i^{S_n^{(i)}},
\end{equation}
where $0\leq p_i$, $\sum_ip_i=1$ and
$\varrho_i^{S_n^{(i)}}$ is a state of the subsystem corresponding to $S_n^{(i)}$ in the split $\alpha_k^{(i)}$
and in this case the $\alpha_k^{(i)}=(S_1^{(i)},S_2^{(i)},\dots,S_k^{(i)})$ $k$-partite splits can be different for different $i$.

Clearly $\mathcal{D}_N^{k+1-\text{sep}}\subset\mathcal{D}_N^{k-\text{sep}}$,
so the notion of $k$-separability gives rise to a natural hierarchic ordering of the states.
The full set of states is $\mathcal{D}_N=\mathcal{D}_N^{1-\text{sep}}$
and we call elements of $\mathcal{D}_N^{k-\text{sep}}\setminus\mathcal{D}_N^{k+1-\text{sep}}$
(i.e.~the $k$-separable but not $k+1$ separable states)
``\emph{$k$-separable entangled}''.
We call the $N$-separable states \emph{fully separable},
the $2$-separable states \emph{biseparable}
and the $1$-separable entangled states \emph{fully entangled}.

Clearly, $\mathcal{D}_N^{\alpha_k}$ is a convex set,
and so is $\mathcal{D}_N^{k-\text{sep}}$, because it is the convex hull of 
the union of $\mathcal{D}_N^{\alpha_k}$s for a given $k$.
Note that these definitions allow  
a $k$-separable state not to be $\alpha_k$-separable for any particular split $\alpha_k$,
and
a state which is $\alpha_k$-separable for all $\alpha_k$ partitions not to be $k+1$-separable.
The existence of such states is somehow conterintuitive,
but explicit examples for these states can be found in literature.
(Using  a method dealing with \emph{Unextendible Product Bases} 
Bennett \textit{et.~al.}~have constructed a three-qubit state
which is separable for all $\alpha_2$ but not fully separable \cite{BennettUPB}.
Another three-qubit example can be found in Ref.~\cite{Mix3QbSLOCC}.)

\begin{figure}
 \setlength{\unitlength}{0.0001652892562\columnwidth}
 \begin{picture}(4840,4840)
  \put(0,0){\includegraphics[width=0.8\columnwidth]{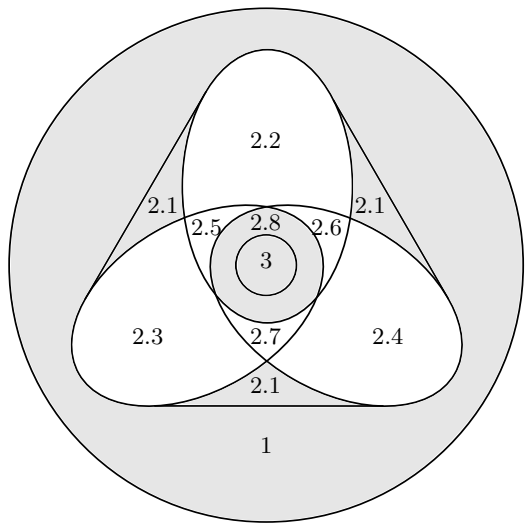}}
  \put(2440,2400){\makebox(0,0)[r]{\strut{}$3$}}
  \put(2440, 700){\makebox(0,0)[r]{\strut{}$1$}}
  \put(2520,1250){\makebox(0,0)[r]{\strut{}$2.1$}}
  \put(2520,2750){\makebox(0,0)[r]{\strut{}$2.8$}}
  \put(2520,1700){\makebox(0,0)[r]{\strut{}$2.7$}}
  \put(2520,3500){\makebox(0,0)[r]{\strut{}$2.2$}}
  \put(1440,1700){\makebox(0,0)[r]{\strut{}$2.3$}}
  \put(3640,1700){\makebox(0,0)[r]{\strut{}$2.4$}}
  \put(3080,2700){\makebox(0,0)[r]{\strut{}$2.6$}}
  \put(1980,2700){\makebox(0,0)[r]{\strut{}$2.5$}}
  \put(3480,2900){\makebox(0,0)[r]{\strut{}$2.1$}}
  \put(1580,2900){\makebox(0,0)[r]{\strut{}$2.1$}}
 \end{picture}
 \caption{Separability classes for three qubits.
The tinted subsets of the diagram can contain permutation-invariant states.}
\label{fig3part}
\end{figure}

Let us now consider the three-qubit case.
(We adopt the notations of \cite{SeevinckUffink}.)
For three qubits we have the partitions: 
$\alpha_1=123$,
$\alpha_2=1-23$,
$\alpha_2=2-31$,
$\alpha_2=3-12$,
$\alpha_3=1-2-3$ (here we use a simplified notation for partitions usual in literature).
With this, the \emph{partial separability classes} of mixed three-qubit states are as follows. 
(See also in Ref.~\cite{SeevinckUffink} and in Fig.~\ref{fig3part}.)

\textit{Class 3:} This is the set of fully separable three-qubit states:
$\mathcal{D}_3^{3-\text{sep}}=\mathcal{D}_3^{1-2-3}$.

\textit{Classes 2.1--2.8:} These are the disjoint subsets of $2$-separable entangled states
$\mathcal{D}_3^{2-\text{sep}}\setminus\mathcal{D}_3^{3-\text{sep}}$.
Classes 2.2--2.8
can be obtained by the set-theoretical intersections of
$\mathcal{D}_3^{1-23}$,
$\mathcal{D}_3^{2-31}$ and
$\mathcal{D}_3^{3-12}$.
(See in Fig.~\ref{fig3part}.)
For example 
\textit{Class 2.8} is $(\mathcal{D}_3^{1-23}\cap\mathcal{D}_3^{2-31}\cap\mathcal{D}_3^{3-12})\setminus \mathcal{D}_3^{1-2-3}$
(i.e.~states that can be writen as convex combination of $3-12$-separable states
and can also be written as convex combination of $2-31$-separable states
and can also be written as convex combination of $1-23$-separable states
but can not be written as convex combination of $1-2-3$-separable states),
\textit{Class 2.7} is $(\mathcal{D}_3^{2-31}\cap\mathcal{D}_3^{3-12})\setminus\mathcal{D}_3^{1-23}$
(i.e.~states that can be written as convex combination of $3-12$-separable states
and can also be written as convex combination of $2-31$-separable states
but can not be written as convex combination of $1-23$-separable states),
\textit{Class 2.2} is $\mathcal{D}_3^{1-23}\setminus(\mathcal{D}_3^{2-31}\cup\mathcal{D}_3^{3-12})$ 
(i.e.~states that can be written as convex combination of $1-23$-separable states
but can not be written as convex combination of $2-31$ or $3-12$-separable states).
On the other hand
the union of the sets of $\alpha_2$-separable states is not a convex one,
it is a genuine subset of its convex hull $\mathcal{D}_3^{2-\text{sep}}$.
This defines \textit{Class 2.1} as $\mathcal{D}_3^{2-\text{sep}}\setminus (\mathcal{D}_3^{1-23}\cup\mathcal{D}_3^{2-31}\cup\mathcal{D}_3^{3-12})$,
i.e.~states that are $2$-separable but can not be written as convex combination of $\alpha_2$-separable states
for any particular $\alpha_2$.
However, we do not consider these states fully entangled
since they can be mixed without the use of genuine three-partite entanglement.

\textit{Class 1:} This contains all the fully entangled states of the system:
$\mathcal{D}_3^{1-\text{sep}}\setminus\mathcal{D}_3^{2-\text{sep}}$.

\section{A symmetric family of mixed three-qubit states}
\label{secRho}

Let $\varrho$ be the mixture of
the Greenberger-Horne-Zeilinger state,
the W state and
the maximally mixed three-qubit state:
\begin{equation}
\label{lo}
\varrho=d\frac{1}{8}\Id
+g\cket{\text{GHZ}}\bra{\text{GHZ}}
+w\cket{\text{W}}\bra{\text{W}},
\end{equation}
where $0\leq d,g,w\leq1$  are real numbers giving rise to the probability distribution characterizing the mixture,
i.e.~$d+g+w=1$.
(In the following sections we plot the subsets of states for which the separability criteria hold on the $g$-$w$-plane,
i.e.~we project the probability-simplex onto the $d=0$ plane.
A point on this plane determines the third coordinate: $d=1-g-w$.
Sometimes it is instructive to use the renormalized parameters 
$\td=d/8$, $\tg=g/2$, $\tw=w/3$.)
In Eq.~(\ref{lo}) $\Id$ denotes the $8\times8$ identity matrix and
the usual GHZ and W states are
\begin{subequations}
\begin{align}
\label{w}
\cket{\text{GHZ}}&=\frac{1}{\sqrt{2}}\Bigl(\cket{000}+\cket{111}\Bigr),\\
\label{ghz}
\cket{\text{W}}&=\frac{1}{\sqrt{3}}\Bigl(\cket{001}+\cket{010}+\cket{100}\Bigr).
\end{align}
\end{subequations}
These two states are representative elements of the two different \emph{SLOCC-classes} \cite{SLOCC} of 
genuine-entangled three-qubit states.~\cite{SLOCCPure3Qb}
The \emph{GHZ state} is maximally entangled in the sense that its one-partite subsystems are maximally mixed.
On the other hand, its two-partite subsystems are separable (having diagonal density matrix).
The one-partite subsystems of the \emph{W-state} are less mixed than the ones of the GHZ state,
but its two-partite subsystems are entangled with Wootters-concurrence $2/3$.
(See Appendix \ref{secWoott}.)

The \emph{GHZ-W mixture} ($d=0$ line) is well studied:
the three tangle \cite{CKWThreetangle} 
with its convex roofs \cite{UhlmannConvexRoof}, 
the Wootters-concurrences \cite{HillWoottersConc,WoottersConc}, 
the one tangle
and the mixed-state CKW-inequality \cite{CKWThreetangle}
were given for this mixture in the paper of Lohmayer \textit{et.~al.}~\cite{MixedThreetangle}.
These results give an upper bound for values of these quantities on the whole simplex defined in Eq.~(\ref{lo}):
if $f(\varrho_{g,w})=\min\sum_ip_if(\psi_i)$ where the minimum is taken over all decomposition 
$\sum_ip_i\cket{\psi_i}\bra{\psi_i}=\varrho_{g,w}$,
and $\varrho_{d,g,w}=d\Id/8+(1-d)\varrho_{g/(1-d),w/(1-d)}$,
and $f(\psi)=0$ on product states
then $f(\varrho_{d,g,w})\leq(1-d)f(\varrho_{g/(1-d),w/(1-d)})$.

The \emph{maximally mixed three-qubit state} can be regarded in some sense 
as the ``center'' of the set of density matrices.
On the other hand this state is sometimes called ``white noise'' because of its uniform spectrum.
Mixing a state with white noise
is the way to investigate the effect of environmental decoherence.~\cite{GuhneTothEntDet}
A noisy state is usually of full rank,
so methods for density matrices of low rank
(like range criterion \cite{HorodeckiRangeCrit},
or finding optimal decompositions with respect some pure-state measures)
usually fail for such states.

On the other hand, there are exact results for the \emph{GHZ-white noise mixture} ($w=0$ line).
In Ref.~\cite{DurCiracMult1} D\"ur and Cirac, 
using their results about a special class of GHZ-diagonal states
have shown 
that $\varrho$ is fully separable \emph{if and only if} $0\leq g\leq1/5$.
Moreover, it follows from their observations that 
if the state is separable under a bipartition then it is fully separable,
so Class 2.8 is empty for these states.
In Ref.~\cite{GuhneSevinckCrit} G\"uhne and Seevinck
gives \emph{necessary and sufficient} condition of genuine three-partite entanglement for GHZ-diagonal states,
which contain the noisy GHZ state:
for $1/5< g\leq3/7$ the state is biseparable, yet inseparable under bipartitions, i.e.~in Class 2.1,
and for $3/7< g\leq1$ the state is fully entangled.
Unfortunately there are no such results for other subsets of the simplex given in Eq.~(\ref{lo}).

The noisy GHZ-W mixture given in Eq.~(\ref{lo}) is clearly a permutation invariant one,
hence the reduced density matrices of $\varrho$ are all of the same form:
$\varrho^{12}=\varrho^{23}=\varrho^{31}$
and 
$\varrho^1=\varrho^2=\varrho^3$,
where $\varrho^{12}=\Tr_3\varrho$, 
$\varrho^1=\Tr_{23}\varrho$ and so forth.
The explicit forms of these matrices are given in Eqs.~(\ref{MxR23}) and (\ref{MxR1}) of the Appendix.

What can we say about the separability-classes of Section \ref{secClasses}
for permutation-invariant three-qubit states \emph{in general}?
Clearly, if a permutation-invariant state is in $\mathcal{D}_3^{\alpha_2}$ for \emph{a particular} $\alpha_2$,
then it is in $\mathcal{D}_3^{\alpha_2}$ for \emph{every} $\alpha_2$.
So permutation-invariant states can not be in Classes 2.2-2.7,
we have to investigate separability criteria only for Class 2.1, Class 2.8 and Class 3.
(Fig.~\ref{fig3part}.)

(Note that the biseparability of a permutation-invariant state
does not mean that the decomposition of Eq.~(\ref{sepk}) contains only permutation-invariant members,
since if the latter holds then the state must be the white noise.
To see this, write out a member of the decomposition 
with the help of the $\sigma_i$ Pauli-matrices and $x_i$, $y_j$ real coefficients
as 
$\varrho^1\otimes\varrho^{23}
=\frac12(\Id+\sum_ix_i\sigma_i)\otimes\frac14(\Id\otimes\Id+\sum_jy_j\sigma_j\otimes\sigma_j)
=\frac18(\Id\otimes\Id\otimes\Id+\sum_ix_i\sigma_i\otimes\Id\otimes\Id+\sum_jy_j\Id\otimes\sigma_j\otimes\sigma_j
+\sum_{ij}x_iy_j\sigma_i\otimes\sigma_j\otimes\sigma_j)$
which can be permutation-invariant if and only if $x_i=0$, $y_j=0$.
The reverse of this is that for permutation-invariant states in Classes 2.1 and 2.8
there does not exist a decomposition as in Eq.~(\ref{sepk}) containing \emph{only} permutation-invariant members.)

The remaining question is wether the remaining classes can contain permutation-invariant states in general. 
Class 1 and Class 3 is clearly nonempty for permutation-invariant states,
and for Classes 2.1 and 2.8 we show explicit examples in Appendix \ref{secExampl}.

If we consider the $2\times2\times2$ three-qubit system 
as a $2\times4$ qubit-qudit system then some well-known and easy-to-use two-partite separability criteria
give rise to separability criteria for $\bigcap_{\alpha_2}\mathcal{D}_3^{\alpha_2}$,
hence for the union of Class 2.8 and Class 3.
(This one is also a convex set since it is the intersection of convex ones.)
First we investigate these criteria.

\section{Two-partite separability criteria}
\label{secTwopart}

In this section we consider our system as a $2\times4$ qubit-qudit system
(with Hilbert-spaces $\mathcal{H}^A=\mathcal{H}^1$ and $\mathcal{H}^B=\mathcal{H}^{23}$)
and investigate some criteria of $1-23$-separability
which means the union of Classes 2.8 and 3.
To do this we will need the spectra of the density matrix $\varrho$ given in Eq.~(\ref{lo}) and its marginals.
(The explicit forms of these matrices are given in Eqs.~(\ref{MxR}), (\ref{MxR23}) and (\ref{MxR1}) of the Appendix.)
Due to the special structure of $\varrho$ finding the eigenvalues of these matrices is not a difficult task.
It turns out that all of the relevant eigenvalues are linear in the parameters $g$ and $w$:
\begin{subequations}
\begin{align}
\label{spectRho}
\Spect(\varrho)&=
\begin{aligned}[t]\bigl\{
\;\td+2\tg &=(3+21g-3w)/24,\\
\td+3\tw   &=(3-3g+21w)/24,\\
\td        &=(3-3g-3w)/24\quad\text{6 times}\bigr\},
\end{aligned}\\
\label{spectRho23}
\Spect(\varrho^{23})&=
\begin{aligned}[t]\bigl\{
\qquad 
2\td+2\tw    &=(6-6g+10w)/24,\\
2\td+\tg+\tw &=(6+6g+2w)/24,\\
2\td+\tg     &=(6+6g-6w)/24,\\
2\td         &=(6-6g-6w)/24 \quad \bigr\},
\end{aligned}\\
\label{spectRho1}
\Spect(\varrho^{1})&=
\begin{aligned}[t]\bigl\{\;
4\td+\tg+2\tw &=(12+4w)/24\\
4\td+\tg+\tw  &=(12-4w)/24 \;\bigr\}.
\end{aligned}
\end{align}
\end{subequations}
Here and in the following, we give expressions with and without $\tilde\;$.
This is because the expressions with the quantities $\td,\tg,\tw$ can be expressive
as they refer to the original mixing weights,
on the other hand we plot in the $g,w$ coordinates.

\subsection{Majorization criterion}
\label{secMaj}

First of all we invoke
the notion of \emph{majorization for probability distributions.}
Let $p=(p_1,\dots,p_n)$ and $q=(q_1,\dots,q_n)$ be two probability distributions of length $n$.
The definition of majorization is as follows.
First we order $p$ and $q$ in non-increasing order 
(we denote this as $p^\downarrow $).
Then $p$ is majorized by $q$ by definition when
the following inequality holds for all $k$:
\begin{equation}
\label{major}
\sum_{i=1}^k p_i^\downarrow \leq \sum_{i=1}^k q_i^\downarrow,  \qquad 1\leq k\leq n.
\end{equation}
This is denoted by $p\prec q$.
(For $k=n$ the inequality turns to equality since both sides of it are equal to $1$.
If the length of $p$ and $q$ differs, one can add some zeroes to the shorter one.)

The majorization is clearly \emph{reflexive} 
($p\prec p$)
and \emph{transitive} 
(if $p\prec q$ and $q\prec r$ then $p\prec r$)
but the \emph{antisymmetry}
(if $p\prec q$ and $q\prec p$ then $p=q$)
holds only in a restricted manner: if $p\prec q$ and $q\prec p$ then $p^\downarrow=q^\downarrow$.
Hence the majorization defines a \emph{partial order} on the set of probability distributions \emph{up to permutations}.
It is clear that $p\nprec q$ does not imply $q\prec p$,
in other words there exist pairs of probability distributions which we can not compare by majorization.
(For example let $p^\downarrow:=(1/2,1/8,\dots)$ and
$q^\downarrow:=(1/3,1/3,\dots)$,
then $p\nprec q$ and $q\nprec p$.)
This is why the majorization gives rise merely to a \emph{partial} ordering.

With respect to majorization
the set of probability distributions contains a greatest and a smallest element.
One can check that all  
probability distribution $p$ majorize the uniform distribution and $p$ is majorized by the distribution containing only one element:
$(\frac{1}{n},\frac{1}{n},\dots)\prec p\prec (1,0,\dots)$.
Employing the notion of majorization we can compare the amount of disorder contained in different probability distributions.
If $p\prec q$ then we can say that $p$ is more disordered than $q$ or equivalently
$q$ is more ordered than $p$,
but there are pairs of distributions for which their rank of disorder can not be compared.

The \emph{majorization of density matrices} is defined via the corresponding majorization of their spectra.
Let $\omega$ and $\sigma$ be two density matrices, then $\omega\prec\sigma$ by definition when
\begin{equation}
\Spect(\omega)\prec\Spect(\sigma).
\end{equation}

Now we can turn to the \emph{majorization criterion} for two-partite systems.
It has been found by Nielssen and Kempe \cite{NielssenKempeMaj}, and it states that
\emph{for a separable state the whole system is more disordered than any of its subsystems:}
\begin{equation}
\label{critMaj}
\varrho\quad\text{separable}\qquad\Longrightarrow\qquad 
\varrho\prec\varrho^A \quad\text{and}\quad \varrho\prec\varrho^B.
\end{equation}
The rhs.~of (\ref{critMaj}) can also be true for entangled states,
but if it does not hold then the state must be entangled.

\begin{figure}
 \centering
 \includegraphics[width=0.8\columnwidth]{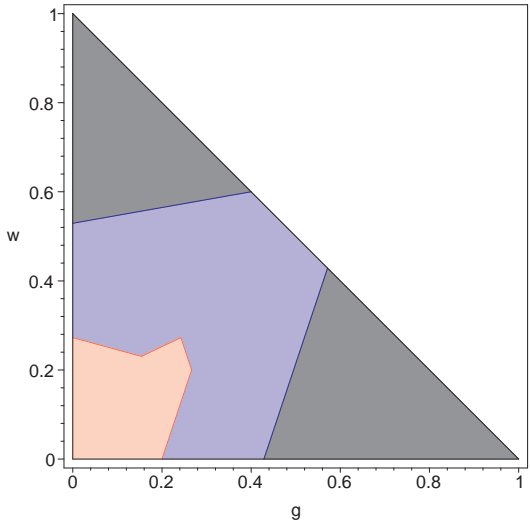}
 \caption{(Color online) Majorization criterion for state (\ref{lo}) on the $g$-$w$-plane. 
(Red/light grey domain: $\varrho\prec\varrho^1$ and $\varrho\prec\varrho^{23}$,
blue/grey domain: $\varrho\prec\varrho^1$ and $\varrho\nprec\varrho^{23}$,
dark grey domains: $\varrho\nprec\varrho^1$ and $\varrho\nprec\varrho^{23}$.)}
 \label{figMaj}
\end{figure}

Let us see what the majorization criterion states about the noisy GHZ-W mixture $\varrho$ given by Eq.~(\ref{lo}).
We can write out the rhs.~of (\ref{critMaj}) explicitly
using the spectra given by Eqs.~(\ref{spectRho})-(\ref{spectRho1}),
then we have to decide whether the inequalities in (\ref{major}) hold.
For this we have to order the eigenvalues of the density matrices in non-increasing order.
These orderings depend on the ranges of the parameters
and it turns out that we have to distinguish between four cases.
These cases are as follows: 
$0\leq g\leq\frac{2}{3}w$,
$\frac{2}{3}w\leq g\leq w$,
$w\leq g\leq\frac{4}{3}w$ and
$\frac{4}{3}w\leq g\leq1$.
It also turns out that in all these cases 
every inequality of (\ref{major}) holds except three ones.
These are as follows:
\begin{widetext}
\begin{equation} 
\label{tablicsku} 
\varrho\in\bigcap_{\alpha_2}\mathcal{D}_3^{\alpha_2}\qquad\Longrightarrow\qquad
\begin{array}{c|cc|c} 
\text{case} & (i) & (ii) & (iii) \\ 
\hline 
0\leq g\leq\frac{2}{3}w & w\leq\frac{3}{11}-\frac{3}{11}g & w\leq 1-3g            & w\leq\frac{9}{17}+\frac{3}{17}g\\ 
\frac{2}{3}w\leq g\leq w& w\leq\frac{3}{19}+\frac{3}{19}g & w\leq 1-3g            & w\leq\frac{9}{17}+\frac{3}{17}g\\ 
w\leq g\leq\frac{4}{3}w & 3g-\frac{3}{5}\leq w            & w\leq 1-3g            & 3g-\frac{9}{7}\leq w \\ 
\frac{4}{3}w\leq g\leq1 & 3g-\frac{3}{5}\leq w  & w\leq\frac{3}{11}-\frac{3}{11}g & 3g-\frac{9}{7}\leq w
\end{array} 
\end{equation}
\end{widetext}
where in columns (\textit{i}) and (\textit{ii}) are the first two inequalities of (\ref{major})
(i.e. $k=1,2$)
written on $\varrho\prec\varrho^{23}$,
and in column (\textit{iii})  are the first inequalities of (\ref{major})
written on $\varrho\prec\varrho^1$ in all of the four cases.
We can make the inequalities of (\ref{tablicsku}) expressive with the help of Fig.~\ref{figMaj}.
It can be seen that in our case $\varrho\prec\varrho^{23}$ implies $\varrho\prec\varrho^1$,
so the bigger subsystem (the trace map on smaller subsystem) gives the stronger condition.
(This is not true in general. One can find a permutation invariant three-qubit state
where $\varrho\prec\varrho^1$ and $\varrho\prec\varrho^{23}$ can hold independently.)

The rhs.~of (\ref{critMaj}) holds for states of parameters in the red/light grey domain of Fig.~\ref{figMaj},
so it contains Classes 2.8 and 3.
On the other hand, states of parameters in the blue/grey or dark grey domain are in Classes 2.1 or 1,
but there can also be such states in the red/light grey domain.
Moreover, \emph{the union of Classes 2.8 and 3 is a convex set inside the red/light grey domain.}
In the following we consider some other criteria in order to decrease the area of the red/light grey domain.
In this way we can identify more states to be in Classes 2.1 or 1.
But before this, we can make an interesting observation here.
One can check that
\emph{for the GHZ-white noise mixture} ($w=0$ line) the majorization criterion 
$\varrho\prec\varrho^1$ and $\varrho\prec\varrho^{23}$
is necessary and sufficient for full-separability,
moreover, the criterion
$\varrho\prec\varrho^1$ and $\varrho\nprec\varrho^{23}$
is necessary and sufficient for Class 2.1,
and the criterion
$\varrho\nprec\varrho^1$ and $\varrho\nprec\varrho^{23}$
is necessary and sufficient for Class 1.
(See Section \ref{secRho} for summary of known exact results on the GHZ-white noise mixture.)
Hence the condition of genuine three-partite entanglement 
is the violation of both majorization of (\ref{critMaj}) for the GHZ-white noise mixture.

\subsection{Entropy criterion}
\label{secEntr}

The \emph{R\'enyi entropy} of a probability distribution $p$ is defined 
for all $0\leq\alpha$ as
\begin{subequations}
\begin{equation}
\label{renyi}
H_\alpha(p)=\frac{1}{1-\alpha}\ln\sum_i p_i^\alpha.
\end{equation}
For $\alpha=0$ this is the logarithm of the number of nonzero $p_i$s, known as \emph{Hartley entropy}:
\begin{equation}
\label{hartley}
H_0(i):=\lim_{\alpha\to0}H_\alpha(p)=\ln|\{p_i\mid p_i\neq0\}|.
\end{equation}
For $\alpha\to1$ it converges to the \emph{Shanon entropy}:
\begin{equation}
\label{shanon}
H_1(p):=\lim_{\alpha\to1}H_\alpha(p)=H(p)=-\sum_i p_i\ln p_i.
\end{equation}
For $\alpha\to\infty$ it converges to the \emph{Chebyshew entropy}:
\begin{equation}
\label{chebyshew}
H_\infty(p):=\lim_{\alpha\to\infty}H_\alpha(p)=-\ln p_\text{max}.
\end{equation}
\end{subequations}
The quantum versions of these are defined on density operators
and can be calculated as the corresponding entropies of the spectrum.
The \emph{quantum-R\'enyi entropy} for all $0\leq\alpha$ is
\begin{subequations}
\begin{equation}
\label{qrenyi}
S_\alpha(\varrho)=\frac{1}{1-\alpha}\ln\Tr(\varrho^\alpha)=H_\alpha(\Spect(\varrho)).
\end{equation}
The \emph{quantum-Hartley entropy} is then:
\begin{equation}
\label{qhartley}
S_0(\varrho):=\lim_{\alpha\to0}S_\alpha(\varrho)=\ln\rk\varrho=H_0(\Spect(\varrho))
\end{equation}
the logarithm of the rank of $\varrho$.
For $\alpha\to1$ it converges to the \emph{von Neumann entropy}:
\begin{equation}
\label{neumann}
\begin{split}
S_1(\varrho):&=\lim_{\alpha\to1}S_\alpha(\varrho)=S(\varrho)=-\Tr\varrho\ln\varrho\\
&=H(\Spect(\varrho)).
\end{split}
\end{equation}
For $\alpha\to\infty$ it converges to the \emph{quantum-Chebyshew entropy}:
\begin{equation}
\label{qchebyshew}
\begin{split}
S_\infty(\varrho):&=\lim_{\alpha\to\infty}S_\alpha(\varrho)=-\ln \max\Spect(\varrho)\\
&=H_\infty(\Spect(\varrho)).
\end{split}
\end{equation}
\end{subequations}

Now we can turn to the \emph{entropy criterion} for two-partite density matrices
\cite{Horodecki2EntCrit,Horodecki3EntCrit,TerhalEntCrit,VollbrechtWolfEntCrit}.
This is an entropy-based restatement of the statement
``for a separable state the whole system is more disordered than its subsystems'':
\begin{multline}
\label{critEntr}
\varrho\quad\text{separable}\qquad\Longrightarrow\qquad \\
S_\alpha(\varrho)\geq S_\alpha(\varrho^A) \quad\text{and}\quad S_\alpha(\varrho)\geq S_\alpha(\varrho^B).
\end{multline}
The rhs.~of (\ref{critEntr}) can also be true for entangled states,
but if it does not hold then the state must be entangled.
The entropy criterion follows from the majorization criterion
since the R\'enyi entropies are \emph{Schur concave} functions on the set of probability distributions.
(That is if $p\prec q$ then $H_\alpha(p)\geq H_\alpha(q)$.)
Therefore \emph{the entropy criterion can not be stronger than the majorization criterion.}
In the following we illustrate this with the state $\varrho$ given in Eq.~(\ref{lo})
for some particular choice of $\alpha$.

The rank of $\varrho$, $\varrho^{23}$ and $\varrho^1$ 
can be determined easily due to the simple form of the spectra in Eqs.~(\ref{spectRho})-(\ref{spectRho1}).
Hence the entropy criterion for Hartley entropy (\ref{qhartley}) can be readily examined.
$\rk\varrho=8$ if and only if $d\neq0$.
The rhs.~of (\ref{critEntr}) holds for these states.
It is true for all states that $\rk\varrho^1=2$.
On the line $w=1-g$ ($d=0$) we have to make distinction between the pure and mixed cases.
If $g=1$ (pure GHZ state) or $w=1$ (pure W state) then $\rk\varrho^{23}=2$ and $\rk\varrho=1$ 
hence for this case $S_0(\varrho)\ngeq S_0(\varrho^1)$ and $S_0(\varrho)\ngeq S_0(\varrho^{23})$.
For the genuine mixtures of GHZ and W states $\rk\varrho^{23}=3$ and $\rk\varrho=2$
hence $S_0(\varrho)\geq S_0(\varrho^1)$ but $S_0(\varrho)\ngeq S_0(\varrho^{23})$.
So we can establish that \emph{the entropy criterion in the limit $\alpha\to0$
(quantum-Hartley entropy)
is too weak, it identifies only the GHZ-W mixture to be entangled.}

\begin{figure}
 \centering
 \includegraphics[width=0.8\columnwidth]{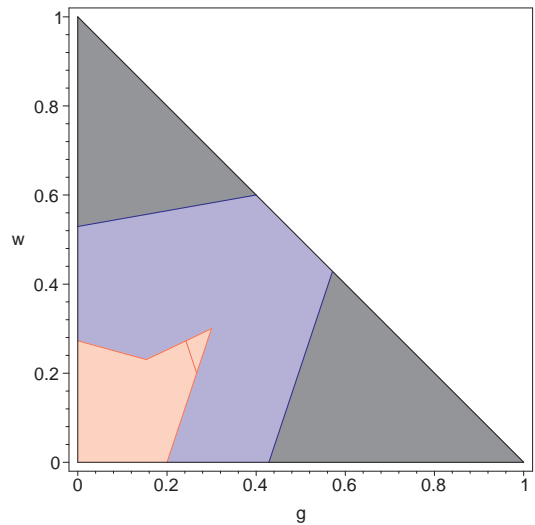}
 \caption{(Color online) Entropy criterion in the limit $\alpha\to\infty$ for state (\ref{lo}) on the $g$-$w$-plane.
(Red/light grey domain: $S_\infty(\varrho) \geq S_\infty(\varrho^1)$ and $S_\infty(\varrho) \geq S_\infty(\varrho^{23})$,
blue/grey domain:       $S_\infty(\varrho) \geq S_\infty(\varrho^1)$ and $S_\infty(\varrho)\ngeq S_\infty(\varrho^{23})$,
dark grey domains:      $S_\infty(\varrho)\ngeq S_\infty(\varrho^1)$ and $S_\infty(\varrho)\ngeq S_\infty(\varrho^{23})$.)}
 \label{figSinf}
\end{figure}

Consider now the entropy criterion in the $\alpha\to\infty$ limit.
This can easily be done because the inequalities of the rhs.~of (\ref{critEntr})
are the same as the ones in the (\textit{i})th and (\textit{iii})th column of (\ref{tablicsku}),
which are written on the maximal eigenvalues.
Hence in this case we have fewer restrictions, 
and one can see in Fig.~\ref{figSinf} that
the rhs.~of (\ref{critEntr}) holds for more states than the rhs.~of (\ref{critMaj})
in the case of the majorization criterion.
Hence \emph{the entropy criterion in the $\alpha\to\infty$ limit 
(quantum-Chebyshew entropy)
identifies a little bit fewer state to be entangled
than the majorization criterion.}

\begin{figure}
 \centering
 \includegraphics[width=0.8\columnwidth]{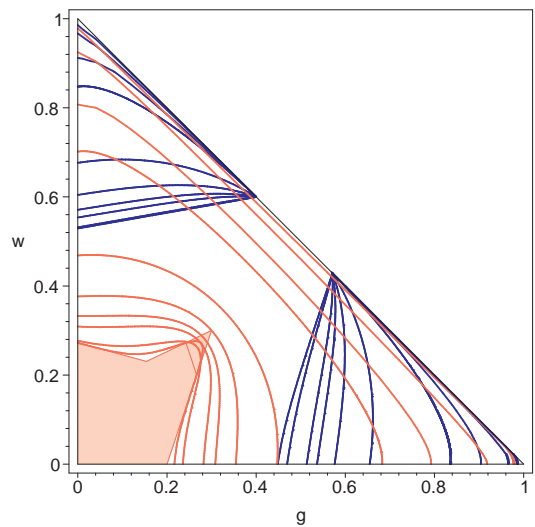}
 \caption{(Color online) Entropy criterion for 
$\alpha=\frac{1}{4}, \frac{2}{4}, \frac{3}{4}, 1, 2, 3, 4, 5, 10, 20$ for state (\ref{lo}) on the $g$-$w$-plane.
(Red/grey  curves: the border of the domain $S_\alpha(\varrho)\geq S_\alpha(\varrho^1)$ and $S_\alpha(\varrho) \geq S_\alpha(\varrho^{23})$,
blue/black curves: the border of the domain $S_\alpha(\varrho)\geq S_\alpha(\varrho^1)$ and $S_\alpha(\varrho)\ngeq S_\alpha(\varrho^{23})$.
Red/light grey domain: copied from Fig.~\ref{figSinf} of the $\alpha\to\infty$ case.)}
 \label{figS}
\end{figure}

Increasing $\alpha$ from $0$ to $\infty$ one can see in Fig.~\ref{figS}
how the borderlines of the domains of the entropy criterion
shrink to the ones in Fig.~\ref{figSinf}.
It is not true in general 
that if $H_\alpha(p)\leq H_\alpha(q)$ and $\alpha\geq\beta$ then $H_\beta(p)\leq H_\beta(q)$.
For these particular spectra it seems that the domains of smaller $\alpha$s would contain the domains of larger $\alpha$s,
but for the large values of $\alpha$ one can see that this is not true.
However, no line can cross the border of the domain of majorization criterion,
since the entropy criterion can not be stronger than the majorization criterion. 

\subsection{Partial transposition criterion}
\label{secPPT}

If a two-partite state is separable
then the partial transposition on subsystem $A$ acts on the $\varrho^A_i$s of the decomposition given in Eq.~(\ref{sep}).
The transposition does not change the eigenvalues of a self-adjoint matrix,
so $(\varrho^A_i)^T$s are also density matrices (i.e.~self-adjoint matrices of trace one).
Hence the partial transpose of a separable density matrix is also a density matrix.
(Its eigenvalues are not the same in general as the ones of the original matrix,
but they are also \emph{nonnegative} and sum up to one.)
The reverse is not true unless the system is of qubit-qubit or qubit-qutrit \cite{PeresHorodeckiCrit},
so our $2\times4$ system is the smallest one for which this implication is only one-way.
In general we get the \emph{partial transposition criterion} of Peres \cite{PeresCrit}:
\begin{equation}
\label{critPPT}
\varrho\quad\text{separable}\qquad\Longrightarrow\qquad \varrho^{T_A} \geq 0.
\end{equation}
It is clear that no matter which subsystem is transposed.

(The partial transposition criterion is the consequence of the \emph{positive maps criterion}: \cite{PeresHorodeckiCrit}
\begin{multline}
\label{critPosMaps}
\varrho\quad\text{separable}\qquad\Longleftrightarrow\qquad\\
(\Phi\otimes\Id)\varrho \geq 0\quad
\text{for all positive maps $\Phi$}.
\end{multline}
This is a necessary and sufficient criterion, but we can not check it for all $\Phi$.
But we can consider a particular class of positive maps to obtain necessary but not sufficient criteria.
For example for $\Phi(\omega)=\omega^T$ we get back the partial transposition criterion.)

\begin{figure}
 \setlength{\unitlength}{0.001904761905\columnwidth}
 \begin{picture}(420,415)
  \put(0,0){\includegraphics[width=0.8\columnwidth]{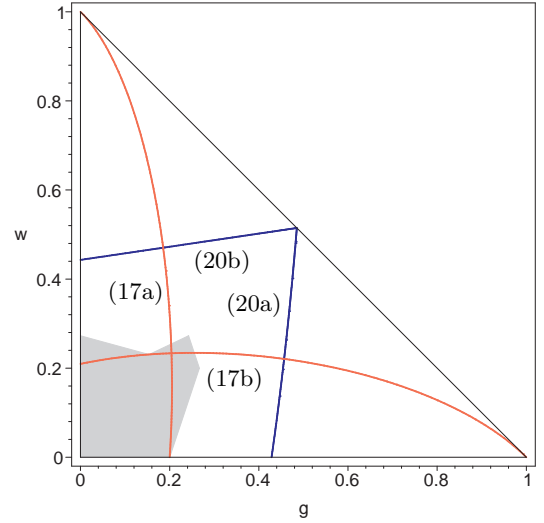}}
  \put(120,180){\makebox(0,0)[r]{\strut{}(\ref{critPPT1})}}
  \put(200,110){\makebox(0,0)[r]{\strut{}(\ref{critPPT2})}}
  \put(215,170){\makebox(0,0)[r]{\strut{}(\ref{critRed3})}}
  \put(190,205){\makebox(0,0)[r]{\strut{}(\ref{critRed4})}}
 \end{picture}
 \caption{(Color online) Partial transposition and reduction criteria for state (\ref{lo}) on the $g$-$w$-plane.
(Eqs.~(\ref{critPPT1}) and (\ref{critPPT2}) of partial transposition criterion hold inside the intersection of the red/grey ellipses.
Blue/black curves: the borders of domains inside the additional Eqs.~(\ref{critRed3}) and (\ref{critRed4}) of reduction criterion hold.
Red/light grey domain: copied from Fig.~\ref{figMaj} of majorization criterion.)}
 \label{figPPTRed}
\end{figure}

Let us apply the partial transposition criterion to the state $\varrho$ of Eq.~(\ref{lo}).
The spectrum of $\varrho^{T_1}$ can easily be calculated due to its block-structure.
(See in Eq.~(\ref{MxRT1}) of the Appendix.)
\begin{multline}
\label{spectRhoPT}
\Spect(\varrho^{T_1})=\\
\begin{aligned}[t]\bigl\{\;
\td+\tw/2\pm&\sqrt{4\tg^2+\tw^2}/2 \\
&=(3-3g+w\pm4\sqrt{9g^2+w^2})/24,\\ 
\td+\tg/2\pm&\sqrt{\tg^2+8\tw^2}/2 \\
&=(3+3g-3w\pm2\sqrt{9g^2+32w^2})/24,\\
\td+\tg &=(3+9g-3w)/24,\\
\td+2\tw &=(3-3g+13w)/24,\\
\td &=(3-3g-3w)/24\quad\text{2 times}\;\bigr\}.
\end{aligned}
\end{multline}
Only the minus-version of the first two eigenvalues can be less than zero
hence we get two inequalities for the positivity of $\varrho^{T_1}$:
\begin{subequations}
\begin{align}
\varrho&\in\bigcap_{\alpha_2}\mathcal{D}_3^{\alpha_2}\qquad\Longrightarrow \notag\\
\label{critPPT1}
&\left\{
\begin{aligned}
0&\leq \td^2+\td\tw-\tg^2\\
0&\leq -135g^2- 15w^2- 6gw-18g+ 6w+9,
\end{aligned}
\right.\\
\label{critPPT2}
&\left\{
\begin{aligned}
0&\leq \td^2+\td\tg-2\tw^2\\
0&\leq - 27g^2-119w^2-18gw+18g-18w+9.
\end{aligned}
\right.
\end{align}
\end{subequations}
Each inequality of these holds inside an ellipse.
These ellipses intersect nontrivially
and in the intersection the rhs.~of (\ref{critPPT}) holds.
(Red/light grey curves in Fig.~\ref{figPPTRed}.)
$g=1/5$ and $w=(24\sqrt2-9)/119=0.209589\dots$ are
the bounds for the union of Class 2.8 and 3 
for the GHZ-white noise ($w=0$) and the W-white noise ($g=0$) mixtures respectively.

The partial transposition criterion states then that
if a state is in Classes 2.8 or 3 then its parameters are inside the intersection of the ellipses,
but there can also be states of Classes 2.1 or 1 in there.
On the other hand the states must be in Classes 2.1 or 1 for parameters outside.
\emph{The inequalities of (\ref{critPPT1}) and (\ref{critPPT2}) are strong in detection of GHZ and W state respectively.}
In Fig.~\ref{figPPTRed} we have also plotted the corresponding domain of the majorization criterion.
(One can check that the only intersection-points of the borderlines 
of the corresponding domains of the two criteria 
are $(g=2/13,w=3/13)$ and $(g=1/5,w=0)$.
This criterion is also a necessary and sufficient one for the 
full separability of the $w=0$ GHZ-white noise mixture.)
It can be seen that 
\emph{the partial transposition criterion 
gives stronger condition than the majorization criterion,}
it identifies more state to be in Classes 2.1 or 1.
Hence the majorization criterion can not identify entangled states of positive partial transpose (PPTES)
on the simplex defined in Eq.~(\ref{lo}).

The PPTESs are exotic entangled states.
They are \emph{bound entangled (undistillable)} states, \cite{Horodecki3BoundEnt} 
that is entangled states from which no entanglement can be distilled at all.
(The entanglement distillation \cite{BennettDistillation,ClarisseThesis} is a family of methods 
which allow one to extract locally maximally entangled pure states out of a given state or its copies.)
It is usually hard to check that a state of positive partial transpose is not separable,
there are few explicit examples of PPTES in the literature
(see a list of references in Section 1.2.4 of Ref.~\cite{ClarisseThesis}).
All the states in Class 2.8 are PPTESs.

\subsection{Reduction criterion}
\label{secRed}

The next one of the examined criteria is the \emph{reduction criterion}. 
\cite{HorodeckiRedCrit,CerfRedCrit}
It states that
\begin{multline}
\label{critRed}
\varrho\quad\text{separable}\qquad\Longrightarrow\qquad \\
\varrho^A\otimes\Id^B-\varrho\geq0 \quad\text{and}\quad \Id^A\otimes\varrho^B-\varrho\geq0.
\end{multline}
This is the consequence of the positive maps criterion given in (\ref{critPosMaps})
for the particular positive map $\Phi(\omega)=\Tr(\omega)\Id - \omega$.
The importance of this criterion is that
its violation is sufficient criterion of \emph{distillability}.~\cite{HorodeckiRedCrit}
It is known \cite{HorodeckiRedCrit} that 
\emph{the reduction criterion can not be stronger than the partial transposition criterion
and they are equivalent for qubit-qudit systems.}
Since our state $\varrho$ defined in Eq.~(\ref{lo}) is the permutation invariant one of three qubits
considered as a $2\times4$ qubit-qudit system,
the equivalence of these two criteria means that 
some kind of pure state entanglement between $1$ and $23$ can be distilled out
from every state of non positive partial transpose.
In other words \emph{in the simplex defined by Eq.~(\ref{lo})
there are no bound entangled $2\times4$ states of non positive partial transpose.}

We can illustrate the equivalence of the partial transposition and reduction criteria.
To do this we have to examine the positivity of the matrices 
$\Id^1\otimes\varrho^{23}-\varrho$ and
$\varrho^1\otimes\Id^{23}-\varrho$. 
(See in Eqs.~(\ref{MxRRI123}) and (\ref{MxRR1I23}) of the Appendix.)
Since $\Tr(\omega)\Id-\omega = (\sigma_2\omega\sigma_2)^T$ for $2\times2$ matrices
(with the Pauli matrix $\sigma_2=\left[\begin{smallmatrix}0&-i\\i&0\end{smallmatrix}\right]$)
it turns out that
\begin{subequations}
\begin{align}
\label{spectRhoI123}
&\Spect(\Id^1\otimes\varrho^{23}-\varrho)=\Spect(\varrho^{T_1}),\\
&\Spect(\varrho^1\otimes\Id^{23}-\varrho)= \notag \\
\label{spectRho1I23}
&\begin{aligned}[t]\bigl\{
3\td+3\tw/2\pm&\sqrt{4\tg^2+\tw^2}/2 \\
&=(9-9g+3w\pm4\sqrt{9g^2+w^2})/24, \\
3\td+\tg\pm\sqrt{2}\tw&=(9+3g-9w\pm8\sqrt{2}w)/24, \\
3\td+\tg+2\tw &=(9+3g+7w)/24\quad\text{2 times}, \\
3\td+\tg+\tw &=(9+3g-w)/24\quad\text{2 times}\;\bigr\}.
\end{aligned}
\end{align}
\end{subequations}
For $\Id^1\otimes\varrho^{23}-\varrho\geq0$ we have the same conditions as in Eqs.~(\ref{critPPT1})-(\ref{critPPT2})
of the partial transposition criterion.
The additional inequalities arise from the minus-version of the first two eigenvalues of $\varrho^1\otimes\Id^{23}-\varrho$:
\begin{subequations}
\begin{align}
\varrho&\in\bigcap_{\alpha_2}\mathcal{D}_3^{\alpha_2}\qquad\Longrightarrow \notag\\
\label{critRed3}
&\left\{
\begin{aligned}
0&\leq 9\td^2-\tg^2+9\td\tw+2\tw^2\\
0&\leq -63g^2- 7w^2- 54gw-162g+ 54w+81,
\end{aligned}
\right.\\
\label{critRed4}
&\left\{
\begin{aligned}
0&\leq 3\td+\tg-\sqrt{2}\tw\\
0&\leq 3g-(9+8\sqrt{2})w+9.
\end{aligned}
\right.
\end{align}
\end{subequations}
The first one of them is true outside a hyperbola, 
the second one is true under a line. (Blue/black curves in Fig.~\ref{figPPTRed}.)

It can be seen that the last two inequalities (\ref{critRed3})-(\ref{critRed4}) do not restrict the 
ones in Eqs.~(\ref{critPPT1})-(\ref{critPPT2}),
as it has to be,
and because of (\ref{spectRhoI123}) 
\emph{the reduction criterion and the partial transposition criterion 
hold for the same states of the GHZ-W-white noise mixture.}
Here we get the stronger condition for the map $\Phi(\omega)=\Tr(\omega)\Id - \omega$ 
acting on the smaller subsystem.
We can also observe that
the inequalities of (\ref{critRed3}) and (\ref{critRed4}) are good in detection of GHZ and W state respectively,
but not so good as the ones of partial transposition criterion.
However, one can check that on the $w=0$ GHZ-white noise mixture
the reduction criterion
$\Id^1\otimes\varrho^{23}-\varrho\geq0$ and $\varrho^1\otimes\Id^{23}-\varrho\geq0$
is necessary and sufficient for full-separability,
the criterion
$\Id^1\otimes\varrho^{23}-\varrho\ngeq0$ and $\varrho^1\otimes\Id^{23}-\varrho\geq0$
is necessary and sufficient for Class 2.1,
and the criterion
$\Id^1\otimes\varrho^{23}-\varrho\ngeq0$ and $\varrho^1\otimes\Id^{23}-\varrho\ngeq0$
is necessary and sufficient for Class 1
in the same fashion as in the majorization criterion of Section \ref{secMaj}.

\subsection{Reshuffling criterion}
\label{secResh}

The \emph{reshuffling criterion} is independent of the partial transposition criterion,
so it can detect entangled states of positive partial transpose.
It states that:
\begin{equation}
\label{critReshuff}
\varrho\quad\text{separable}\qquad\Longrightarrow\qquad 
\Vert R(\varrho) \Vert_\text{Tr} \leq 1,
\end{equation}
where the \emph{trace-norm} is $\Vert A \Vert_\text{Tr}=\Tr\sqrt{A^\dagger A}$,
and the \emph{reshuffling map} $R$ is defined on matrix elements as
$[R(\varrho)]_{ii',jj'}=\varrho_{ij,i'j'}$.

\begin{figure}
 \setlength{\unitlength}{0.001904761905\columnwidth}
 \begin{picture}(420,415)
  \put(0,0){\includegraphics[width=0.8\columnwidth]{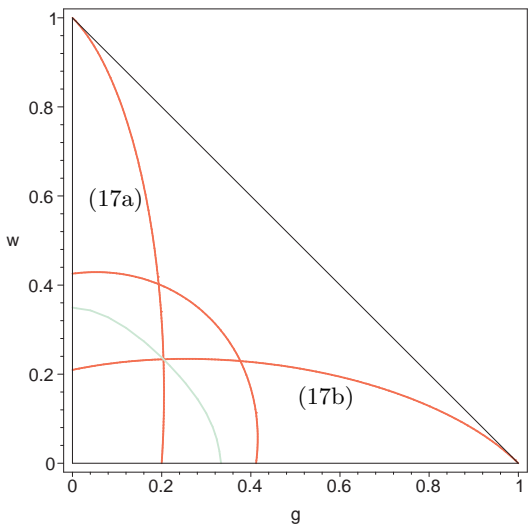}}
  \put(110,260){\makebox(0,0)[r]{\strut{}(\ref{critPPT1})}}
  \put(280,100){\makebox(0,0)[r]{\strut{}(\ref{critPPT2})}}
 \end{picture}
 \caption{(Color online) Reshuffling criteria for state (\ref{lo}) on the $g$-$w$-plane.
(Red/grey curve: reshuffling criterion for $2\times4$ system as in Section \ref{secResh},
green/light grey curve: reshuffling criterion for $2\times2\times2$ system as in Section \ref{secPerm}.
We have also copied the borderlines of the domains in which (\ref{critPPT1}) and (\ref{critPPT2}) of the partial transposition criterion hold 
from Fig.~\ref{figPPTRed}.
The inequalities hold on the side of the curves containing the origin.)}
 \label{figResh}
\end{figure}

The four nonzero singular values of the $4\times16$ reshuffled density matrix, 
(see in Eq.~(\ref{MxRR24}) of the Appendix)
i.e.~the square root of the nonnegative eigenvalues of $R(\varrho)R(\varrho)^\dagger$ are:
\begin{multline}
\Spect\sqrt{R(\varrho)R(\varrho)^\dagger} \\
=\left\{
\sqrt{p_1\pm2\sqrt{p_2}}/2,
\sqrt{\tg^2+2\tw^2},
\sqrt{\tg^2+2\tw^2}
\right\},
\end{multline}
where
\begin{equation}
\begin{split}
p_1=& 16\td^2 +4\tg^2 +10\tw^2 +8\td\tg +12\td\tw,\\
p_2=& 64\td^4 +9\tw^4 +64\td^3\tg +96\td^3\tw +12\td\tw^3 \\
& +16\td^2\tg^2 +40\td^2\tw^2 +4\tg^2\tw^2 \\
& +80\td^2\tg\tw +16\td\tg^2\tw +24\td\tg\tw^2.
\end{split}\notag
\end{equation}
The sum of them is less or equal than $1$ inside a curve of high degree
which can be seen in Fig.~\ref{figResh} (red/grey curve).
States of Classes 2.8 and 3 must be inside this curve,
states outside this curve must belong to Classes 2.1 or 1,
but one can see that \emph{this criterion does not restrict the partial transposition criterion,}
it does not detect PPTESs
in the GHZ-W-white noise mixture of Eq.~(\ref{lo}).

\section{Three-partite separability criteria}
\label{secThreepart}

In this section we consider our system given in Eq.~(\ref{lo})
as a proper $2\times2\times2$ three-qubit one
and investigate some general $3$-qubit $k$-separability criteria.

\subsection{Permutation criterion}
\label{secPerm}

First consider the \emph{permutation criterion} in general \cite{HorodeckiPerm}.
Note that the reshuffling and the partial transpose of a density matrix are
nothing else than the permutation of the local matrix indices.
Moreover, since
the trace norm is the sum of the absolute values of the eigenvalues for hermitian matrices
and the trace is invariant under partial transposition
it turns out that $\varrho^{T_1}\geq0$ if and only if $\Vert \varrho^{T_1} \Vert_\text{Tr}=1$.
So the partial transposition criterion (\ref{critPPT})
and the reshuffling criterion (\ref{critReshuff}) can be formulated in the same fashion.
Moreover, this can be done for $N$ subsystems in a general way. \cite{HorodeckiPerm}
Let $\pi\in S_{2N}$ a permutation of the $2N$ matrix indices
and let $\Lambda_\pi$ the map realizing this index permutation:
if $\varrho=\sum \varrho_{i_1i_2\dots i_N,i_{N+1}i_{N+2}\dots i_{2N}}\cket{i_1i_2\dots i_N}\bra{i_{N+1}i_{N+2}\dots i_{2N}}$
then
$[\Lambda_\pi(\varrho)]_{i_{\pi(1)}i_{\pi(2)}\dots i_{\pi(N)},i_{\pi(N+1)}i_{\pi(N+2)}\dots i_{\pi(2N)}}
=\varrho_{i_1i_2\dots i_N,i_{N+1}i_{N+2}\dots i_{2N}}$.
Now the \emph{permutation criterion} states that
\begin{multline}
\label{critPerm}
\varrho\quad\text{fully separable}\qquad\Longrightarrow\qquad\\ 
\Vert \Lambda_\pi(\varrho) \Vert_\text{Tr} \leq 1,\quad \forall \pi\in S_{2N}.
\end{multline}

The permutation criterion gives $(2N)!$ criteria
but not all of them are inequivalent.
It is well known \cite{HorodeckiPerm} that 
\emph{for two subsystems, every criteria given by the permutation criterion turn out to be equivalent
either the partial transposition criterion or the reshuffling criterion.}
In Ref.~\cite{ClarissePerm} Clarisse has shown that there are only six inequivalent criteria
in the case of three subsystems: 
three one-partite-transpositions and three two-partite-reshufflings.
For our \emph{permutation-invariant} three-qubit system
all the one-partite-transpositions give the same condition
which we have already investigated in Section \ref{secPPT}.
On the other hand, all the two-partite-reshufflings give another condition
which is a new one.

So let $R'=\Lambda_\pi$ the map implementing the reshuffling of the $2$ and $3$ subsystems:
 $[R'(\varrho)]_{ijj',i'kk'}=\varrho_{ijk,i'j'k'}$.
(See in Eq.~(\ref{MxRR222}) of the Appendix.)
With this we have to calculate the eigenvalues of the matrix
$R'(\varrho)R'(\varrho)^\dagger$ for the two-parameter state $\varrho$ given in Eq.~(\ref{lo}).
This $8\times8$ matrix can be transformed
(by simultaneous row-column permutation)
into blockdiagonal form
consisting three blocks of the shape $3\times3$, $3\times3$ and $2\times2$.
However, the forms of the $g,w$-depending eigenvalues of the $3\times3$ blocks are still too complicated,
so we only plot the border of the domain in which the criterion (\ref{critPerm}) holds.
(See green/light grey curve in Fig.~\ref{figResh}.)

The condition $\Vert R'(\varrho) \Vert_\text{Tr} \leq 1$ holds inside the green/light grey curve
in Fig.~\ref{figResh}.
This figure suggests that
this reshuffling does not give stricter condition for full separability
than the partial transposition criterion, hence it can not identify PPTESs.
However, we can not be sure in this
 due to the difficult computation of $\Vert R'(\varrho) \Vert_\text{Tr}$.
Fully separable states must be enclosed by the curves belonging to (\ref{critPPT1})-(\ref{critPPT2}) of partial transposition criterion,
states outside this domain must belong to Classes 2.8, 2.1 or 1.
However, in Section \ref{secPPT} the partial transposition criterion has yielded condition for Classes 2.8 and 3,
so we can conclude that states outside this domain must belong to Classes 2.1 or 1,
the Class 2.8 is completely restricted into this domain.

\subsection{Criteria on spin-observables}
\label{secSpin}

In Ref.~\cite{SeevinckUffink} Seevinck and Uffink introduced a systematic way
to obtain necessary criteria of separability for all the separability-classes of an $N$-qubit system.
Their criteria generalize some previously known criteria, (see Refs.~in Ref.~\cite{SeevinckUffink})
such as Laskowski-\.Zukowski criterion (necessary for $k$-separability),
Mermin-type separability inequalities (necessary for $k$-separability),
Fidelity-criterion (necessary for $2$-separability)
and D\"ur-Cirac depolarization criterion (necessary for $\alpha_k$-separability).
We consider the three-qubit case and get criteria for Class 2.1, Class 2.8 and Class 3
given in Section \ref{secClasses}.

The method of Seevinck and Uffink deals with \emph{three orthogonal spin-observables} on each subsystem: 
$(X^{(1)},Y^{(1)},Z^{(1)})$. 
Here the superscript $(1)$ denotes that these are single-qubit operators.
Let $I^{(1)}$ denote the $2\times2$ identity matrix.
From the $(X^{(1)},Y^{(1)},Z^{(1)},I^{(1)})$ one-qubit observables acting on the $2$ and $3$ subsystem
one can form
two sets of two-qubit observables: 
$(X_x^{(2)},Y_x^{(2)},Z_x^{(2)},I_x^{(2)})$. 
Here the superscript $(2)$ denotes that these are two-qubit operators
and $x=0,1$ refers to the two sets:
\begin{align}
\label{spin2}
X_0^{(2)}&=\frac12\left(X^{(1)}\otimes X^{(1)} - Y^{(1)}\otimes Y^{(1)}\right),\notag\\
X_1^{(2)}&=\frac12\left(X^{(1)}\otimes X^{(1)} + Y^{(1)}\otimes Y^{(1)}\right),\notag\\
Y_0^{(2)}&=\frac12\left(Y^{(1)}\otimes X^{(1)} + X^{(1)}\otimes Y^{(1)}\right),\notag\\
Y_1^{(2)}&=\frac12\left(Y^{(1)}\otimes X^{(1)} - X^{(1)}\otimes Y^{(1)}\right),\notag\\
Z_0^{(2)}&=\frac12\left(Z^{(1)}\otimes I^{(1)} + I^{(1)}\otimes Z^{(1)}\right),\notag\\
Z_1^{(2)}&=\frac12\left(Z^{(1)}\otimes I^{(1)} - I^{(1)}\otimes Z^{(1)}\right),\notag\\
I_0^{(2)}&=\frac12\left(I^{(1)}\otimes I^{(1)} + Z^{(1)}\otimes Z^{(1)}\right),\notag\\
I_1^{(2)}&=\frac12\left(I^{(1)}\otimes I^{(1)} - Z^{(1)}\otimes Z^{(1)}\right).
\end{align}
(Note that $I_x^{(2)}$s are \emph{not} identity operators.)
From this two-qubit observables 
and the one-qubit ones acting on the $1$ subsystem
one can form
four sets of three-qubit observables acting on the full system:
$(X_x^{(3)},Y_x^{(3)},Z_x^{(3)},I_x^{(3)})$. 
Here the superscript $(3)$ denotes that these are three-qubit operators
and $x=0,1,2,3$ refers to the four sets: 
\begin{align}
\label{spin3}
X_y^{(3)}    &=\frac12\left(X^{(1)}\otimes X^{(2)}_{y/2} - Y^{(1)}\otimes Y^{(2)}_{y/2}\right),\notag\\
X_{y+1}^{(3)}&=\frac12\left(X^{(1)}\otimes X^{(2)}_{y/2} + Y^{(1)}\otimes Y^{(2)}_{y/2}\right),\notag\\
Y_y^{(3)}    &=\frac12\left(Y^{(1)}\otimes X^{(2)}_{y/2} + X^{(1)}\otimes Y^{(2)}_{y/2}\right),\notag\\
Y_{y+1}^{(3)}&=\frac12\left(Y^{(1)}\otimes X^{(2)}_{y/2} - X^{(1)}\otimes Y^{(2)}_{y/2}\right),\notag\\
Z_y^{(3)}    &=\frac12\left(Z^{(1)}\otimes I^{(2)}_{y/2} + I^{(1)}\otimes Z^{(2)}_{y/2}\right),\notag\\
Z_{y+1}^{(3)}&=\frac12\left(Z^{(1)}\otimes I^{(2)}_{y/2} - I^{(1)}\otimes Z^{(2)}_{y/2}\right),\notag\\
I_y^{(3)}    &=\frac12\left(I^{(1)}\otimes I^{(2)}_{y/2} + Z^{(1)}\otimes Z^{(2)}_{y/2}\right),\notag\\
I_{y+1}^{(3)}&=\frac12\left(I^{(1)}\otimes I^{(2)}_{y/2} - Z^{(1)}\otimes Z^{(2)}_{y/2}\right),
\end{align}
for $y=0,2$.
(Note that $I_x^{(3)}$s are \emph{not} identity operators.)

Now for particular $\alpha_2$ investigating some relations among the expectation-values of these operators
with respect to the state $\varrho$
one can get some nontrivial inequalities
valid for all $\varrho\in\mathcal{D}_3^{\alpha_2}$.
From these one can form inequalities valid for a given separability class of Section \ref{secClasses}.
Here we recall \cite{SeevinckUffink} these criteria for the classes we need to deal with:
\begin{align}
\label{critSU2}
\varrho\in&\mathcal{D}_3^{2-\text{sep}}\qquad\Longrightarrow\notag\\
&\sqrt{\bracket{X_x^{(3)}}^2+\bracket{Y_x^{(3)}}^2}\leq\sum_{y\neq x}\sqrt{\bracket{I_y^{(3)}}^2-\bracket{Z_y^{(3)}}^2}
\end{align}
for $x=0,1,2,3$,
\begin{align}
\label{critSU283}
\varrho\in&\bigcap_{\alpha_2}\mathcal{D}_3^{\alpha_2} \qquad\Longrightarrow\notag\\
&\begin{aligned}
\max_x\Bigl\{\bracket{X_x^{(3)}}^2+&\bracket{Y_x^{(3)}}^2\Bigr\}\\
\leq \min_x\Bigl\{&\bracket{I_x^{(3)}}^2-\bracket{Z_x^{(3)}}^2\Bigr\}\leq1/4
\end{aligned}
\end{align}
and
\begin{align}
\label{critSU3}
\varrho\in&\mathcal{D}_3^{3-\text{sep}}\qquad\Longrightarrow\notag\\
&\begin{aligned}
\max_x\Bigl\{\bracket{X_x^{(3)}}^2+&\bracket{Y_x^{(3)}}^2\Bigr\}\\
\leq\min_x\Bigl\{&\bracket{I_x^{(3)}}^2-\bracket{Z_x^{(3)}}^2\Bigr\}\leq1/16.
\end{aligned}
\end{align}
One has to do some optimisation of the local spin observables $(X^{(1)},Y^{(1)},Z^{(1)})$
to get violation of the respective inequality for a given state.

In the following we will consider some special 
measurement-settings when the observables 
$(X^{(1)},Y^{(1)},Z^{(1)})$ are the same for each subsystem.
Writing out explicitly $(X_x^{(3)},Y_x^{(3)},Z_x^{(3)},I_x^{(3)})$,
one can see that \emph{for a permutation-invariant state}
the squares of the expectation values are the same for $x=1,2,3$, i.e.:
$\bracket{X_1^{(3)}}^2=\bracket{X_2^{(3)}}^2= \bracket{X_3^{(3)}}^2 $,
and the same for $Y_x^{(3)}$s, $Z_x^{(3)}$s and $I_x^{(3)}$s.
Hence we have to consider merely the $x=0,1$ indices.

\begin{figure}
 \setlength{\unitlength}{0.001904761905\columnwidth}
 \begin{picture}(420,415)
  \put(0,0){\includegraphics[width=0.8\columnwidth]{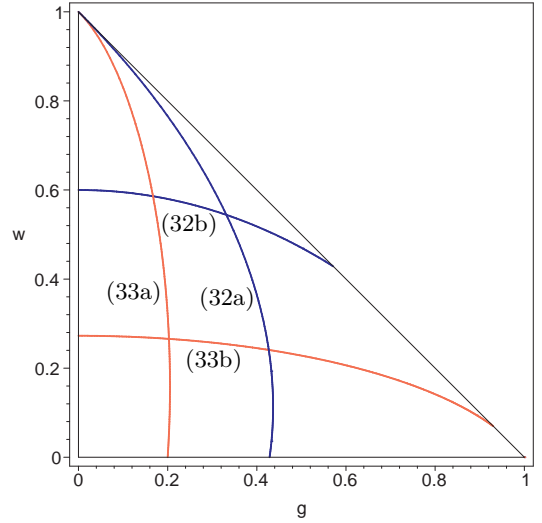}}
  \put(120,180){\makebox(0,0)[r]{\strut{}(\ref{critSU283_1})}}
  \put(185,125){\makebox(0,0)[r]{\strut{}(\ref{critSU283_2})}}
  \put(195,175){\makebox(0,0)[r]{\strut{}(\ref{critSU2_1})}}
  \put(165,235){\makebox(0,0)[r]{\strut{}(\ref{critSU2_2})}}
 \end{picture}
 \caption{(Color online) Criteria on spin-observables for state (\ref{lo}) on the $g$-$w$-plane.
(Red/grey curves: the border of domains inside Eqs.~(\ref{critSU2_1}) and (\ref{critSU2_2}) hold,
blue/black curves: the border of domains inside Eqs.~(\ref{critSU283_1}) and (\ref{critSU283_2}) hold.
The inequalities hold on the side of the curves containing the origin.)}
 \label{figSpin}
\end{figure}

First consider the special choice when
$(X^{(1)},Y^{(1)},Z^{(1)})=(\sigma_x,\sigma_y,\sigma_z)$ for each subsystem.
(Here $\sigma_i$s denote the usual Pauli matrices.)
We refer to this as Setting I.
The inequalities (\ref{critSU2})-(\ref{critSU3}) can be written 
as relatively simple expressions in the matrix elements \cite{SeevinckUffink}:
\begin{align}
\label{critSU2_1g}
\varrho\in&\mathcal{D}_3^{2-\text{sep}}\qquad\Longrightarrow\notag\\
&\begin{aligned}[t]
|\varrho_{07}|&\leq \sqrt{\varrho_{66}\varrho_{11}}+\sqrt{\varrho_{55}\varrho_{22}}+\sqrt{\varrho_{33}\varrho_{44}},\\
|\varrho_{61}|&\leq \sqrt{\varrho_{00}\varrho_{77}}+\sqrt{\varrho_{55}\varrho_{22}}+\sqrt{\varrho_{33}\varrho_{44}},\\
|\varrho_{52}|&\leq \sqrt{\varrho_{66}\varrho_{11}}+\sqrt{\varrho_{00}\varrho_{77}}+\sqrt{\varrho_{33}\varrho_{44}},\\
|\varrho_{34}|&\leq \sqrt{\varrho_{66}\varrho_{11}}+\sqrt{\varrho_{55}\varrho_{22}}+\sqrt{\varrho_{00}\varrho_{77}},
\end{aligned}
\end{align}
\begin{align}
\label{critSU283_1g}
\varrho&\in\bigcap_{\alpha_2}\mathcal{D}_3^{\alpha_2}\qquad\Longrightarrow\notag\\ 
&\begin{aligned}
\max&\left\{|\varrho_{07}|^2,|\varrho_{61}|^2,|\varrho_{52}|^2,|\varrho_{34}|^2\right\}\\
\leq&\min\left\{\varrho_{00}\varrho_{77},\varrho_{66}\varrho_{11},\varrho_{55}\varrho_{22},\varrho_{33}\varrho_{44}\right\}
\leq1/16
\end{aligned}
\end{align}
and
\begin{align}
\label{critSU3_1g}
\varrho&\in\mathcal{D}_3^{3-\text{sep}}\qquad\Longrightarrow\notag\\
&\begin{aligned}
\max&\left\{|\varrho_{07}|^2,|\varrho_{61}|^2,|\varrho_{52}|^2,|\varrho_{34}|^2\right\}\\
\leq&\min\left\{\varrho_{00}\varrho_{77},\varrho_{66}\varrho_{11},\varrho_{55}\varrho_{22},\varrho_{33}\varrho_{44}\right\}
\leq1/64.
\end{aligned}
\end{align}
(Here the matrix indices run from $0$ to $7$ so as to be equal to the binary indices we use later.)
Let us consider another two special measurement settings:
Setting II: $(X^{(1)},Y^{(1)},Z^{(1)})=(\sigma_y,\sigma_z,\sigma_x)$ for each subsystem,
Setting III:  $(X^{(1)},Y^{(1)},Z^{(1)})=(\sigma_z,\sigma_x,\sigma_y)$ for each subsystem.
The inequalities of (\ref{critSU2})-(\ref{critSU3}) written for these two settings
are much more complicated expressions in symbolic matrix elements
than the ones in (\ref{critSU2_1g})-(\ref{critSU3_1g}).
But for the state $\varrho$ given in Eq.~(\ref{lo}) 
it is not too difficoult to write out these inequalities explicitly.
It turns out that for each of these three settings
the $x=1$ inequality of (\ref{critSU2}),
the second inequality of (\ref{critSU283}) and
the second inequality of (\ref{critSU3})
hold for all the parameter values of the simplex.
Because of this, the criteria hold for Class 3
are not stricter than the ones for the union of Class 2.8 and Class 3.
The remaining inequalities for the three measurement settings are as follows:
\begin{subequations}
\begin{align}
\varrho&\in\mathcal{D}_3^{2-\text{sep}}\qquad\Longrightarrow\notag\\
\label{critSU2_1}
\text{I.}&\left\{
\begin{aligned}
\tg&\leq3\sqrt{\td(\td+\tw)}\\
0&\leq -7g^2 -6gw -15w^2 -18g +6w +9,
\end{aligned}
\right.\\
\label{critSU2_2}
\text{II.}&\left\{
\begin{aligned}
3\tw&\leq\sqrt{(8\td+\tw)(8\td+4\tg+\tw)}\\
0&\leq -9g^2 -5w^2 -12w+9,
\end{aligned}
\right.\\
\label{critSU2_3}
\text{III.}&\left\{
\begin{aligned}
\sqrt{4\tg^2+81\tw^2}&\leq3(8\td+2\tg+\tw)\\
0&\leq - g^2 -5w^2 -12w+9,
\end{aligned}
\right.
\end{align}
\end{subequations}
and
\begin{subequations}
\begin{align}
\varrho&\in\bigcap_{\alpha_2}\mathcal{D}_3^{\alpha_2}\qquad\Longrightarrow\notag\\
\label{critSU283_1}
\text{I.}&\left\{
\begin{aligned}
\tg^2&\leq\td(\td+\tw)\\
0&\leq -45g^2 -2gw -5w^2 -6g +2w +3,
\end{aligned}
\right.\\
\label{critSU283_2}
\text{II.}&\left\{
\begin{aligned}
81\tw^2&\leq(8\td+\tw)(8\td+4\tg+\tw)\\
0&\leq -9g^2 -77w^2 -12w+9,
\end{aligned}
\right.\\
\label{critSU283_3}
\text{III.}&\left\{
\begin{aligned}
4\tg^2+81\tw^2&\leq(8\td+2\tg+\tw)^2\\
0&\leq -9g^2 -77w^2 -12w+9,
\end{aligned}
\right.
\end{align}
\end{subequations}
Clearly, the inequality of (\ref{critSU2_3}) is weaker than the one of (\ref{critSU2_2}),
the inequality of (\ref{critSU283_3}) is the same as the one of (\ref{critSU283_2}).
Moreover,
    the inequality of (\ref{critSU283_1}) is the same as the one of (\ref{critPPT1}) of partial transposition criterion,
but the inequality of (\ref{critSU283_2}) is strictly weaker than the other one of partial transposition criterion.
\emph{So these settings does not give stricter conditions for Classes 2.8 and 3 than partial transposition criterion,}
however, we get criteria for biseparability for the first time.
In Fig.~\ref{figSpin} we show the borderlines of the domains of the criteria belonging to Settings I.~and II.
These inequalities restrict Classes 2.1, 2.8 and 3 to be inside the domain enclosed by the blue/black curves
and Classes 2.8 and 3 to be inside the domain enclosed by the red/grey curves.
\emph{We can conclude that Settings I.~and II.~are strong in detection of GHZ and W state respectively.}
One can check that for the $w=0$ GHZ-white noise mixture
the inequalities of (\ref{critSU2_1}) and (\ref{critSU283_1}) of Setting I.~hold 
if and only if the state is fully separable,
(\ref{critSU2_1}) is violated but (\ref{critSU283_1}) holds
if and only if the state is in Class 2.1
and
both of them are violated
if and only if the state is fully entangled.
For the $g=0$ W-white noise mixture
if $3/11<w$ then $\varrho$ is in Class 2.1 or Class 1,
and if $3/5<w$ then $\varrho$ is fully entangled.

However, there are infinitely many criteria depending on the measurement settings
and we do not have a method to find a set of settings leading to the strichtest criterion.
We have tried some other randomly chosen settings
which can be used to reduce the area where the criteria hold.
We could not find settings that give stronger criteria on the $w=0$ or $g=0$ axes of the simplex than Settings I.~and II.~respectively.
We have found settings that excludes states from the corresponding classes,
but these states are far from these axes,
and we have not found settings which give stronger condition for Classes 2.8 and 3 than the partial transposition criterion.
We have found settings by which the condition for biseparability can be strengthened,
but these conditions are just a little bit stronger far from the axes than the ones in Section \ref{secMatrix}.

\subsection{Criteria on matrix elements}
\label{secHub}

In a recent paper \cite{HuberkCrit},
Gabriel \textit{et.~al.}~have given criterion for $k$-separability,
based on their previously derived framework for the detection of biseparability \cite{HuberCrit}.
It turns out that for the noisy GHZ-W mixture given in Eq.~(\ref{lo})
these criteria give the same results as the ones of Seevinck and Uffink, given in the previous section,
but these criteria have the advantage that they can be used \emph{in the same form} not only for qubits,
but for subsystems of arbitrary, even different dimensions.
As our knowledge, these are the only such criteria of $k$-separability.

Consider some permutation operators acting on $\mathcal{H}\otimes\mathcal{H}$,
i.e.~on the two copies of the $N$-partite Hilbert space $\mathcal{H}=\mathcal{H}^1\otimes\mathcal{H}^2\otimes\dots\otimes\mathcal{H}^N$.
Let $P_a$s the operators which swap the $a$th subsystems of the two copies:
$P_a\cket{i_1i_2\dots i_N}\otimes\cket{j_1j_2\dots j_N}=
\cket{i_1i_2\dots i_{a-1}j_ai_{a+1} \dots i_N}\otimes\cket{j_1j_2\dots j_{a-1}i_aj_{a+1} \dots j_N}$
where $\{\cket{i_a}\}$ is basis in $\mathcal{H}^a$.
Now for a composite subsystem $\mathcal{H}^S=\otimes_{a\in S}\mathcal{H}^a$ let $P_S=\prod_{a\in S}P_a$.
The key fact is that 
if the state of a set of subsystems can be separated from the rest of the state
then the corresponding $P_S$ leaves the two copies of the state invariant: $P_S^\dagger\varrho^{\otimes2} P_S=\varrho^{\otimes2}$.
With this and convexity arguments, one can get the following criteria for $k$-separability \cite{HuberkCrit}:
\begin{multline}
\label{critHub}
\varrho\in\mathcal{D}_N^{k-\text{sep}}\qquad\Longrightarrow\qquad \\
\sqrt{\bra{\Phi}\varrho^{\otimes2} P_\text{tot}\cket{\Phi}}
\leq\sum_i\left(\prod_{n=1}^k\bra{\Phi}P^\dagger_{S_n^{(i)}}\varrho^{\otimes2}P_{S_n^{(i)}}\cket{\Phi}\right)^{\frac1{2k}},
\end{multline}
where $\cket{\Phi}\in\mathcal{H}\otimes\mathcal{H}$ is a \emph{fully separable} vector,
and the total swap operator is $P_\text{tot}=\prod_{a=1}^NP_a$.
Here $i$ runs over all posible $k$-partite splits $\alpha_k^{(i)}=(S_1^{(i)},S_2^{(i)},\dots,S_k^{(i)})$.

The inequality in (\ref{critHub}) is written on the matrix elements of $\varrho$
determined by the separable detection-vector $\cket{\Phi}$.
For a given state, optimisation on $\cket{\Phi}$ 
is needed to achieve the violation of (\ref{critHub}).

To apply these criteria to the noisy GHZ-W mixture given in Eq.~(\ref{lo})
we have to choose a suitable detection-vector $\cket{\Phi}$.
It turns out that 
$\cket{\Phi_{\text{GHZ}}}=\cket{000111}$ and $\cket{\Phi_{\text{W}}}=H^{\otimes6}\cket{\Phi_{\text{GHZ}}}$
are good choices for states in the vicinity of GHZ and W states respectively,
as observed in Ref.~\cite{HuberkCrit}.
(Here $H=1/\sqrt2\left[\begin{smallmatrix}1&1\\1&-1\end{smallmatrix}\right]$ 
is matrix of the usual Hadamard/discrete Fourier transformation for qubits.)
With these two vectors we get the same criteria 
for $2$-separability as the ones in (\ref{critSU2_1})   and (\ref{critSU2_2})  respectively,
and 
for $3$-separability as the ones in (\ref{critSU283_1}) and (\ref{critSU283_2})  respectively.
However, (\ref{critSU283_1}) and (\ref{critSU283_2}) obtained by the criteria on spin observables
are criteria not only for Class 3, but for the union of Classes 2.8 and 3,
so in this sense the criteria on spin observables are a bit stronger.

We can not be sure that the detection-vectors above give
the strongest conditions at least for the noisy GHZ and noisy W states.
However, it is an interesting observation
that the Hadamard transformation relates 
not only the two ``optimal'' detection-vectors $\cket{\Phi_{\text{GHZ}}}$ and $\cket{\Phi_{\text{W}}}$
but also the two ``optimal'' measurement-settings Setting I.~$(\sigma_x,\sigma_y,\sigma_z)$ and 
$(\sigma_z,-\sigma_y,\sigma_x)$, (by the transformation $\sigma_i\mapsto H\sigma_iH^\dagger$)
which is equivalent to Setting II.
(This equivalence holds only for \emph{permutation-invariant} three-qubit states,
when the three sets of observables $(X^{(1)},Y^{(1)},Z^{(1)})$ are the \emph{same for each subsystem.}
In this case one can check 
that the quantities $\bracket{X_x^{(3)}}^2+\bracket{Y_x^{(3)}}^2$ for $x=0,1,2,3$
are invariant under the 
transformation
$(X^{(1)},Y^{(1)},Z^{(1)}) \leftrightarrow (Y^{(1)},X^{(1)},Z^{(1)})$
and
$(X^{(1)},Y^{(1)},Z^{(1)}) \leftrightarrow (X^{(1)},-Y^{(1)},Z^{(1)})$.
These can be seen by writing out the definitions given in Eqs.~(\ref{spin3}).)

We have tried some other randomly chosen detection-vectors
which can be used to reduce the area where the criteria hold,
and we get the same observations as at the end of the previous section:
One can strenghtend the conditions only far from the $w=0$ or $g=0$ axes of the simplex,
we have not found detection-vectors which give stronger condition for full-separability than the partial transposition criterion,
and we have found settings by which the condition for biseparability can be strengthened,
but these conditions are just a little bit stronger far from the axes than the ones in Section \ref{secMatrix}.

\subsection{Criteria on matrix elements -- a different approach}
\label{secMatrix}

In Ref.~\cite{GuhneSevinckCrit} G\"uhne and Seevinck have given
some further biseparability and full-separability criteria
on the matrix elements:
\begin{subequations}
\begin{align}
\label{critGS2a}
\varrho\in\mathcal{D}_3^{2-\text{sep}}\qquad\Longrightarrow&\notag\\ 
|\varrho_{07}|&\leq
\sqrt{\varrho_{66}\varrho_{11}}+
\sqrt{\varrho_{55}\varrho_{22}}+
\sqrt{\varrho_{33}\varrho_{44}},\\
\label{critGS2b}
\begin{split}
|\varrho_{12}|+|\varrho_{14}|+|\varrho_{24}|&\leq
\sqrt{\varrho_{00}\varrho_{33}}+
\sqrt{\varrho_{00}\varrho_{55}}+
\sqrt{\varrho_{00}\varrho_{66}}\\
&\quad+\left(\varrho_{11}+\varrho_{22}+\varrho_{44}\right)/2.
\end{split}
\end{align}
\end{subequations}
The criterion (\ref{critGS2a}) is necessary and sufficient for GHZ-diagonal states
and can also be obtained as a special case
of the criteria of Section \ref{secSpin} (Eq.~(\ref{critSU2_1g})).
However, this criterion---and the others in this Section---arises from a quite different approach
as the one in (\ref{critSU2_1g}),
since these criteria
have been derived from direct investigation of the matrix elements of pure separable states
with the use of convexity argument.
The criterion in (\ref{critGS2b}) is independent of the first one
and it is quite strong in detection of W state mixed with white noise.

Of course these and the following inequalities
can be written on local unitary transformed density matrices,
and optimisation with regard to local unitaries might be necessary,
but this can lead to very complicated expressions in the original matrix elements.
An advantage of the method of the previous section is that
it handles the matrix indices through the detection vector $\cket{\Phi}$.

The full-separability criteria of \cite{GuhneSevinckCrit}:
\begin{subequations}
\begin{align}
\label{critGS3_1g}
\varrho\in\mathcal{D}_3^{3-\text{sep}}\qquad\Longrightarrow&\notag\\
|\varrho_{07}|&\leq
\left(\varrho_{11}\varrho_{22}\varrho_{33}\varrho_{44}\varrho_{55}\varrho_{66}\right)^{1/6},\\
\label{critGS3_2g}
|\varrho_{12}|+|\varrho_{14}|+|\varrho_{24}|&\leq
\sqrt{\varrho_{00}\varrho_{33}}+
\sqrt{\varrho_{00}\varrho_{55}}+
\sqrt{\varrho_{00}\varrho_{66}}.
\end{align}
\end{subequations}
(\ref{critGS3_1g}) is
necessary and sufficient for GHZ state mixed with white noise,
and (\ref{critGS3_2g}) is violated in the vicinity of the W state.
Moreover, one can obtain other conditions from (\ref{critGS3_1g}) by making substitutions as follows.
Consider a fully separable pure state $\varrho^{3-\text{sep}}=\cket{\Psi}\bra{\Psi}$, where
$\cket{\Psi}=(a_0\cket{0}+a_1\cket{1})\otimes(b_0\cket{0}+b_1\cket{1})\otimes(c_0\cket{0}+c_1\cket{1})$.
Then the diagonal elements are $(\varrho^{3-\text{sep}})_{ijk,ijk}=|a_i|^2|b_j|^2|c_k|^2$,
where we use the $ijk=000,001,\dots,111$ binary indexing.
Then
\begin{equation*} 
\begin{split}
(\varrho^{3-\text{sep}})_{ijk,ijk}&(\varrho^{3-\text{sep}})_{i'j'k',i'j'k'}\\
&=(\varrho^{3-\text{sep}})_{ijk',ijk'}(\varrho^{3-\text{sep}})_{i'j'k,i'j'k}\\
&=(\varrho^{3-\text{sep}})_{ij'k,ij'k}(\varrho^{3-\text{sep}})_{i'jk',i'jk'}\\
&=(\varrho^{3-\text{sep}})_{i'jk,i'jk}(\varrho^{3-\text{sep}})_{ij'k',ij'k'}.
\end{split}
\end{equation*}
Moreover, 
the rhs.~of the inequality of (\ref{critGS3_1g})
can be written as
$\left(\varrho_{11}^2\varrho_{22}^2\varrho_{33}^2\varrho_{44}^2\varrho_{55}^2\varrho_{66}^2\right)^{1/12}$,
and with these substitutions
we can obtain a third power of four matrix elements under the $1/12$th power.
So we can get expressions of four matrix elements
(e.g.~$\left( \varrho_{11}\varrho_{22}\varrho_{44}\varrho_{77}\right)^{1/4}$)
on the rhs.~of (\ref{critGS3_1g}).
With the substitutions above one can get $28$ different inequalities for (\ref{critGS3_1g}) 
with an expresison of sixth order under the sixth root on the rhs.~and $12$ different ones with an expresison of fourth order under the fourth root.
For permutation-invariant states we have
$\varrho_{11}=\varrho_{22}=\varrho_{44}$ and
$\varrho_{66}=\varrho_{55}=\varrho_{33}$
and for our case $\varrho_{00}=\varrho_{77}$ as well,
hence the number of different inequalities reduces to $8$ and $5$ respectively.
The rhs's of inequality (\ref{critGS3_1g}) which are different for permutation-invariant matrices 
are as follows:
\begin{align*}
\left( \varrho_{00}^3\varrho_{77}^3 \right)^{1/6},
&\left( \varrho_{11}  \varrho_{22}  \varrho_{33}  \varrho_{44}  \varrho_{55}  \varrho_{66}  \right)^{1/6},\\
\left( \varrho_{00}  \varrho_{11}  \varrho_{22}  \varrho_{44}  \varrho_{77}^2\right)^{1/6},
&\left( \varrho_{00}^2\varrho_{33}  \varrho_{55}  \varrho_{66}  \varrho_{77}  \right)^{1/6},\\
\left( \varrho_{00}  \varrho_{11}^2\varrho_{66}^2\varrho_{77}  \right)^{1/6},
&\left( \varrho_{00}^2\varrho_{11}  \varrho_{66}  \varrho_{77}^2\right)^{1/6},\\
\left( \varrho_{11}^2\varrho_{22}  \varrho_{44}  \varrho_{66}  \varrho_{77}\right)^{1/6},
&\left( \varrho_{00}  \varrho_{11}  \varrho_{33}  \varrho_{55}  \varrho_{66}^2\right)^{1/6}
\end{align*}
and
\begin{align*}
\left( \varrho_{00}^2\varrho_{77}^2 \right)^{1/4},&\\
\left( \varrho_{22}  \varrho_{33}  \varrho_{44}  \varrho_{55}   \right)^{1/4},&
\left( \varrho_{00}  \varrho_{11}  \varrho_{66}  \varrho_{77}   \right)^{1/4},\\
\left( \varrho_{11}  \varrho_{22}  \varrho_{44}  \varrho_{77}   \right)^{1/4},&
\left( \varrho_{00}  \varrho_{33}  \varrho_{55}  \varrho_{66}   \right)^{1/4}.
\end{align*}
It turns out that the strongest conditions can be given with the last of these
and with the original one in (\ref{critGS3_1g}).
(We could also make some substitutions 
in the rhs.~of (\ref{critGS3_2g}) but these would not give stronger conditions than the original one.)

\begin{figure}
 \setlength{\unitlength}{0.001904761905\columnwidth}
 \begin{picture}(420,415)
  \put(0,0){\includegraphics[width=0.8\columnwidth]{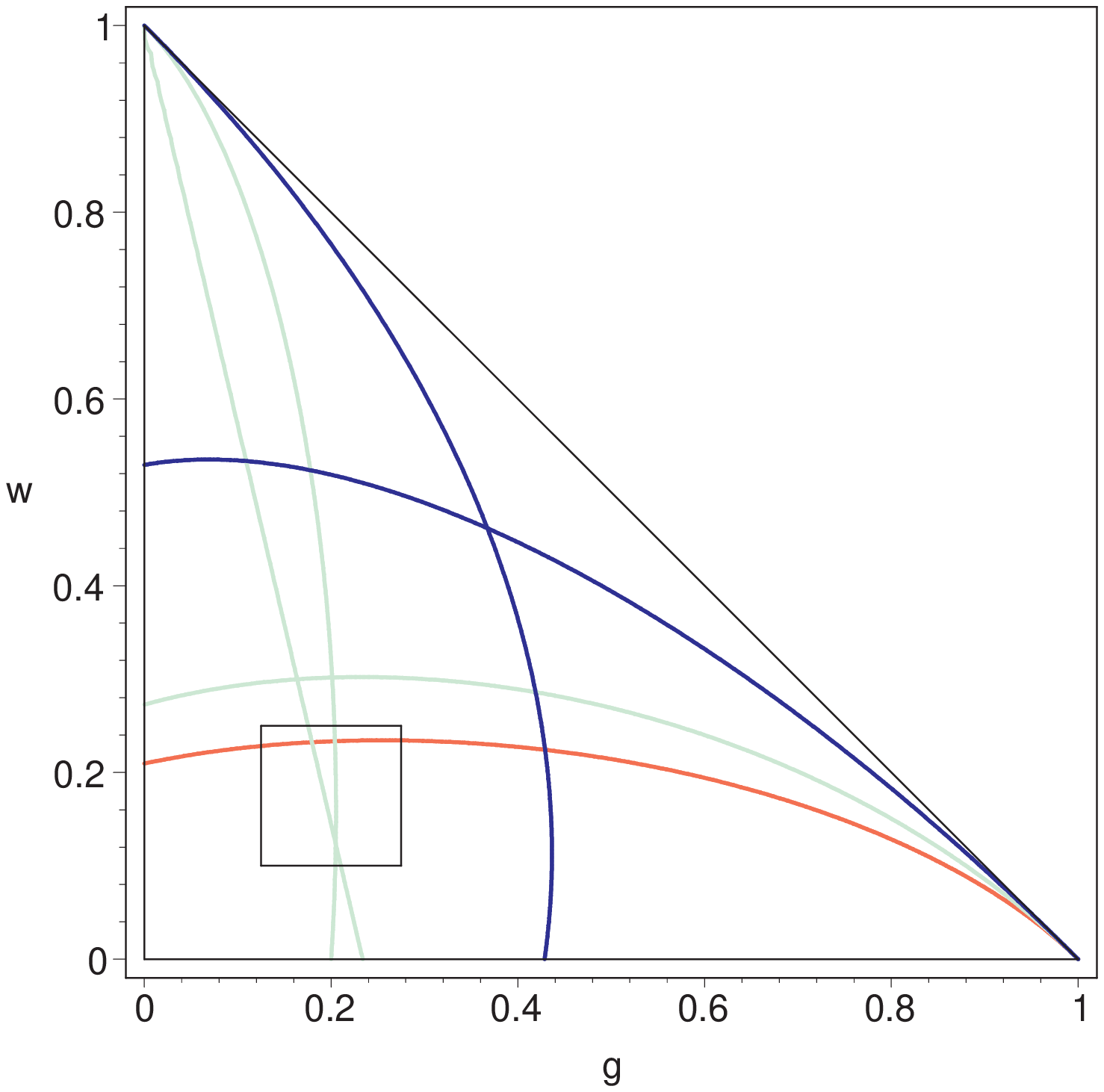}}
  \put(206,208){\includegraphics[width=0.4\columnwidth]{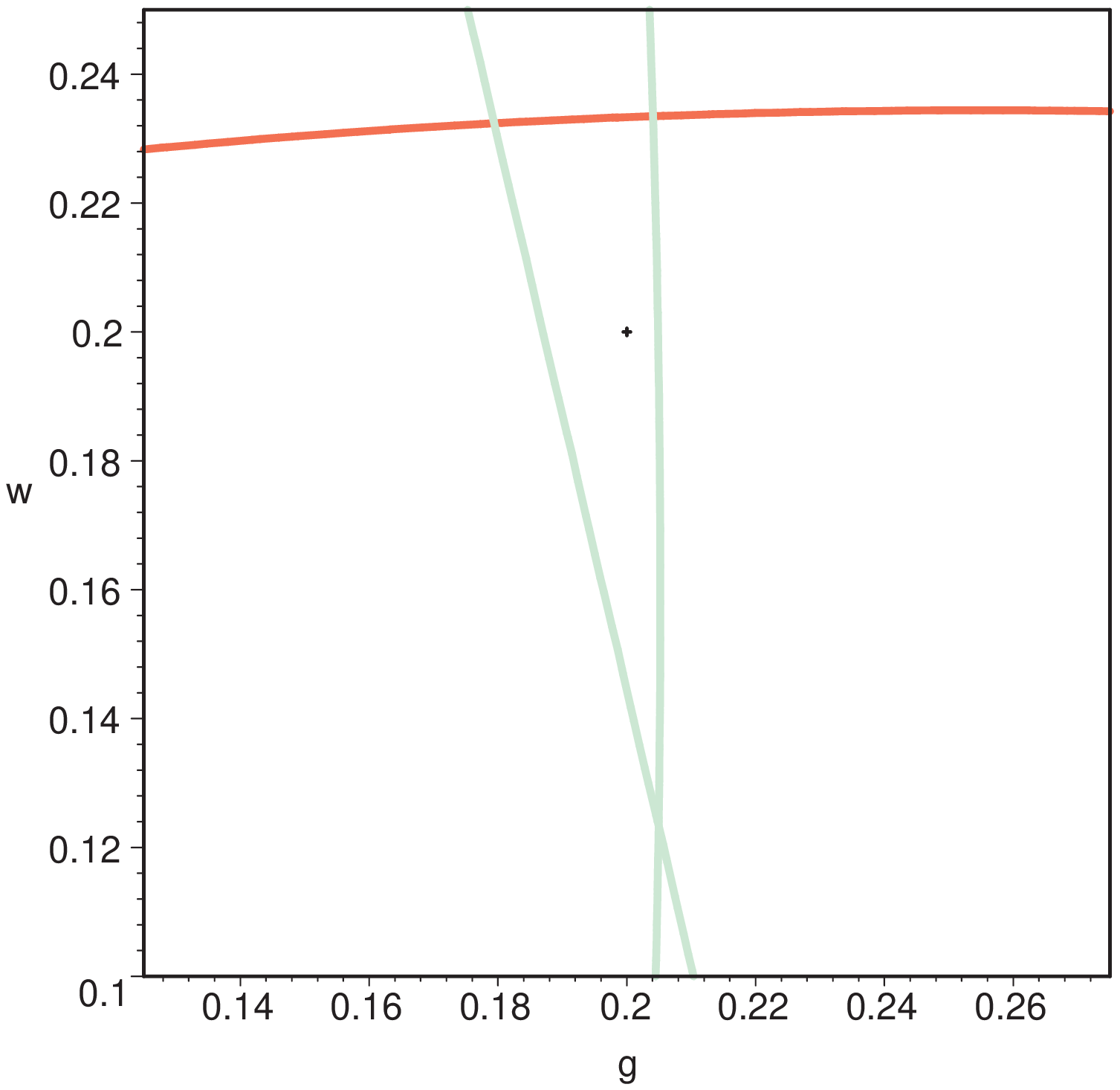}}
  \put(110,280){\makebox(0,0)[r]{\strut{}(\ref{critGS3_1})}}
  \put(100,200){\makebox(0,0)[r]{\strut{}(\ref{critGS3_11})}}
  \put(265,130){\makebox(0,0)[r]{\strut{}(\ref{critGS3_2})}}
  \put(275,100){\makebox(0,0)[r]{\strut{}(\ref{critPPT2})}}
  \put(120,310){\makebox(0,0)[r]{\strut{}(\ref{critGS2_1})}}
  \put(260,160){\makebox(0,0)[r]{\strut{}(\ref{critGS2_2})}}
 \end{picture}
 \caption{(Color online) Criteria on matrix elements for state (\ref{lo}) on the $g$-$w$-plane.
(Green/light grey curves: 
the borders of domains inside Eqs.~(\ref{critGS3_1})-(\ref{critGS3_2}) hold,
blue/black curves: 
the borders of domains inside Eqs.~(\ref{critGS2_1})-(\ref{critGS2_2}) hold.
Red/grey curve: 
the border of domain inside Eq.~(\ref{critPPT2}) of partial transposition criterion hold,
copied from Fig.~\ref{figPPTRed}.
The point $g=1/5$, $w=1/5$ is also shown.
The inequalities hold on the side of the curves containing the origin.)}
 \label{figMxe}
\end{figure}

Writing out the criteria of biseparability and full separability we get:
\begin{subequations}
\begin{align}
\label{critGS2_1}
\varrho&\in\mathcal{D}_3^{2-\text{sep}}\quad\Longrightarrow& 
\tg&\leq 3\sqrt{\td(\td+\tw)} ,\\
\label{critGS2_2}
&&\tw&\leq \sqrt{(\td+\tg)\td}+(\td+\tw)/2
\end{align}
\end{subequations}
and
\begin{subequations}
\begin{align}
\label{critGS3_1}
\varrho&\in\mathcal{D}_3^{3-\text{sep}}\quad\Longrightarrow&\quad 
  \tg&\leq ((\td+\tw)\td)^{1/2},\\
\label{critGS3_11}
&&\tg&\leq ((\td+\tg)\td^3)^{1/4},\\
\label{critGS3_2}
&&\tw&\leq \sqrt{(\td+\tg)\td}.
\end{align}
\end{subequations}
(See in Eq.~(\ref{MxR}) of the Appendix.)
Clearly, the biseparability condition of (\ref{critGS2_1}) is the same as the one of (\ref{critSU2_1}) of the criterion on spin-observables
but condition of (\ref{critGS2_2}) is strictly stronger than the one of (\ref{critSU2_2}).
(On the $g=0$ noisy W state it gives bound $9/17$.)
The full-separability condition of (\ref{critGS3_1}) is the same as the one of (\ref{critSU283_1}) of the criterion on spin-observables
(and the one of (\ref{critPPT1}) of partial transposition criterion as well)
but the condition of (\ref{critGS3_2}) is weaker than 
    the one of (\ref{critSU283_2}) of the criterion on spin-observables.
Hence at this point these criteria are 
stronger for biseparability but weaker for full separability
than the criteria on spin-observables for our state.
But we have another full-separability condition: (\ref{critGS3_11}) 
can be stronger in a region than the ones based on the partial transposition criterion.
The states of parameters in this region are entangled ones of positive partial transpose,
no pure state entanglement can be distilled from them.
The borders of the domains in which these conditions hold 
and the region of PPTESs can be seen in Fig.~\ref{figMxe}.

Now we show a representing matrix of the region of PPTESs determined by (\ref{critPPT1}), (\ref{critPPT2})
and (\ref{critGS3_11}).
One can check that the state of parameters $g=1/5$, $w=1/5$ is contained by this set
and the explicit form of (\ref{lo}) for this point is
\begin{equation} 
\label{PPTESlo}
\varrho^{\text{PPTES}}=\frac1{120}\begin{bmatrix}
 21  &\cdot&\cdot&\cdot&\cdot&\cdot&\cdot& 12  \\
\cdot& 17  &  8  &\cdot&  8  &\cdot&\cdot&\cdot\\
\cdot&  8  & 17  &\cdot&  8  &\cdot&\cdot&\cdot\\
\cdot&\cdot&\cdot&  9  &\cdot&\cdot&\cdot&\cdot\\
\cdot&  8  &  8  &\cdot& 17  &\cdot&\cdot&\cdot\\
\cdot&\cdot&\cdot&\cdot&\cdot&  9  &\cdot&\cdot\\
\cdot&\cdot&\cdot&\cdot&\cdot&\cdot&  9  &\cdot\\
 12  &\cdot&\cdot&\cdot&\cdot&\cdot&\cdot& 21  \\
\end{bmatrix}.
\end{equation}

\section{SLOCC classes of genuine three-qubit entanglement}
\label{secClass1}

A fully entangled three-qubit \emph{pure} state can be either of GHZ-type or of W-type \cite{SLOCCPure3Qb}
in the sense of \emph{Stochastic Local Operations and Classical Communications} (SLOCC, \cite{SLOCC}):
vectors of these two different types can not be transformed into each other by local invertible operations.
These fully entangled vectors $\cket{\psi}$ can be classified by the so called \emph{three-tangle} $\tau_{123}(\psi)$
\cite{CKWThreetangle}
as $\tau_{123}(\psi)\neq0$ exactly for the GHZ-type vectors.

In Ref.~\cite{Mix3QbSLOCC} Ac\'in \textit{et.~al.}~have investigated
the classification of \emph{mixed} three-qubit states in the sense of SLOCC
and they have shown that Class 1 of fully entangled states
can be divided into two subsets, namely the ones of GHZ and W-type entanglement, by the following definitions.
A state is of \emph{W-type} ($\mathcal{D}_3^{\text{W}}$) if it can be expressed as a mixture of projectors onto $2$-separable and W-type vectors
(therefore $\mathcal{D}_3^{\text{W}}$ is also a convex set)
and \emph{GHZ-type} vector is required for a GHZ-type mixed state.
Hence the following holds:
\begin{equation}
\label{incl}
\mathcal{D}_3^{3-\text{sep}}\subset
\mathcal{D}_3^{2-\text{sep}}\subset
\mathcal{D}_3^{\text{W}}\subset
\mathcal{D}_3^{\text{GHZ}}\equiv
\mathcal{D}_3^{1-\text{sep}}\equiv
\mathcal{D}_3.
\end{equation}
Let \emph{Class W} the set $\mathcal{D}_3^{\text{W}}\setminus\mathcal{D}_3^{2-\text{sep}}$
and \emph{Class GHZ} the set $\mathcal{D}_3^{\text{GHZ}}\setminus\mathcal{D}_3^{\text{W}}$,
so $\text{Class 1} = \text{Class W}\cup\text{Class GHZ}$.
The mixed state extension of $\tau_{123}$
\begin{equation} 
\label{mixtau}
\tau_{123}(\varrho)=\min\Bigg\{\sum_ip_i\tau_{123}(\psi_i)\Bigg| \sum_ip_i\cket{\psi_i}\bra{\psi_i}=\varrho\Bigg\}
\end{equation}
is a good indicator for Class GHZ: $\tau_{123}(\varrho)\neq0$ exactly for Class GHZ.

\begin{figure}
 \setlength{\unitlength}{0.001904761905\columnwidth}
 \begin{picture}(420,415)
  \put(0,0){\includegraphics[width=0.8\columnwidth]{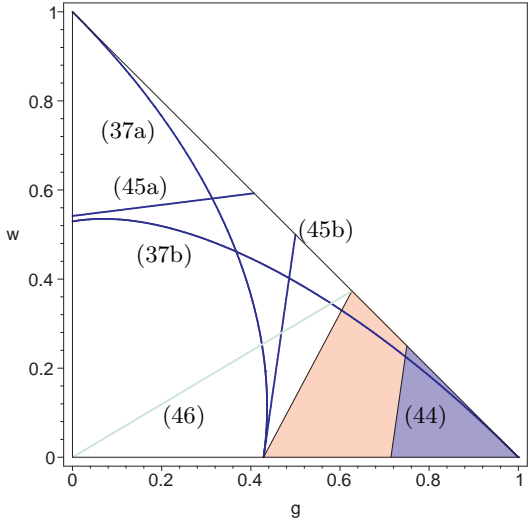}}
  \put(130,265){\makebox(0,0)[r]{\strut{}(\ref{critWW1})}}
  \put(280,230){\makebox(0,0)[r]{\strut{}(\ref{critWW2})}}
  \put(355,80){\makebox(0,0)[r]{\strut{}(\ref{critWGHZ})}}
  \put(160,80){\makebox(0,0)[r]{\strut{}(\ref{critGHZ2})}}
  \put(120,310){\makebox(0,0)[r]{\strut{}(\ref{critGS2_1})}}
  \put(150,210){\makebox(0,0)[r]{\strut{}(\ref{critGS2_2})}}
 \end{picture}
 \caption{(Color online) Criteria on three-partite entanglement classes for state (\ref{lo}) on the $g$-$w$-plane.
(Blue/black straight lines: the borders of domains inside Eqs.~(\ref{critWW1})-(\ref{critWW2}) hold,
blue/black curves of second order: the borders of domains inside Eqs.~(\ref{critGS2_1})-(\ref{critGS2_2}) hold,
copied from Fig.~\ref{figMxe}.
Eq.~(\ref{critWGHZ}) holds inside the blue/grey domain,
and the border of Class GHZ is inside the red/light grey domain.
The inequalities hold on the side of the curves containing the origin.
Eq.~(\ref{critGHZ2}) holds under the green/light grey line.)}
 \label{figWGHZ}
\end{figure}

A method to determine to which class a given mixed state belongs is
the use of witness operators.~\cite{PeresHorodeckiCrit} 
A hermitian operator $W$ is a \emph{witness operator} for a convex compact set $\mathcal{C}\subset\mathcal{D}_N$
if $0\leq\Tr W\sigma$ for all $\sigma\in\mathcal{C}$
and there exists $\varrho\notin\mathcal{C}$ for which $\Tr W\varrho<0$.
Hence the negativity of expectation value of the observable $W$
bears witness that the state does not belong to $\mathcal{C}$.
In Ref.~\cite{Mix3QbSLOCC} there have been given some witnesses 
for $\mathcal{D}_3^{\text{W}}$ and $\mathcal{D}_3^{\text{GHZ}}$:
\begin{equation}
W_{\text{GHZ}}=\frac34\Id-\cket{\text{GHZ}}\bra{\text{GHZ}}
\end{equation}
can detect $\mathcal{D}_3^{\text{GHZ}}$ and
\begin{subequations}
\begin{align}
W_{\text{W}_1}&=\frac23\Id-\cket{\text{W}}\bra{\text{W}},\\
W_{\text{W}_2}&=\frac12\Id-\cket{\text{GHZ}}\bra{\text{GHZ}}
\end{align}
\end{subequations}
can detect $\mathcal{D}_3^{\text{W}}$.
With these we have
\begin{align}
\varrho\in \mathcal{D}_3^{\text{W}}\qquad\Longrightarrow\qquad&\notag\\
\label{critWGHZ}
\begin{split}
0\leq\Tr W_{\text{GHZ}}\varrho&=(20\td-2\tg+9\tw)/4\\
 &=(5-7g+w)/8
\end{split}
\end{align}
and if the inequality is violated then $\varrho\in\text{Class GHZ}$,
as well as
\begin{subequations}
\begin{align}
\varrho\in \mathcal{D}_3^{2-\text{sep}}\qquad\Longrightarrow\qquad&\notag\\
\label{critWW1}
\begin{split}
0\leq\Tr W_{\text{W}_1}\varrho&=(13\td+4\tg-3\tw)/3\\
 &=(13+3g-21w)/24,
\end{split}\\
\label{critWW2}
\begin{split}
0\leq\Tr W_{\text{W}_2}\varrho&=(6\td-2\tg+3\tw)/2\\
 &=(3-7g+w)/8
\end{split}
\end{align}
\end{subequations}
and if either or both of the inequalities is violated then $\varrho\in\text{Class 1}$.
In Fig.~\ref{figWGHZ} we plot the lines on which these inequalities are saturated.
It can be checked that (\ref{critWW1}) and (\ref{critWW2}) gives
weaker condition for biseparability than (\ref{critGS2_1}) and (\ref{critGS2_2}) of the previous Section.
We can conclude that all the states in the blue/grey domain belong to Class GHZ,
and the biseparable states are enclosed by the blue/black curves,
however, both type of fully entangled states can be here too.

The equality in (\ref{critWGHZ}) gives an ``upper bound'' for the border of Class GHZ. 
(See blue/grey domain in Fig.~\ref{figWGHZ}.)
Fortunately, we have a possibility to give also a ``lower bound'' for that,
thanks to the results of Lohmayer \textit{et.~al.}~\cite{MixedThreetangle}.
They have studied the GHZ-W mixture ($d=0$) and 
they have found that there exists a decomposition of projectors onto vectors of vanishing three-tangle
if and only if $0\leq g\leq g_0=4\cdot2^{1/3}/(3+4\cdot2^{1/3})=0.626851\dots$,
hence for these parameters the mixed state extension (\ref{mixtau}) of the three-tangle is zero.
If we mix the states of this interval with white noise
then the three-tangle remains zero and neither of these states can belong to Class GHZ.
So we can state that
\begin{equation}
\label{critGHZ2}
\varrho\in\mathcal{D}_3^{\text{GHZ}}\qquad\Longrightarrow\qquad
w<\frac3{4\cdot2^{1/3}}g.
\end{equation}
which holds under the green/light grey line of Fig.~\ref{figWGHZ}.
This condition is quite weak, but we can make it stronger.
Recall that on the $w=0$ line (noisy GHZ state)
$\varrho\in\text{Class 1}$ if and only if $3/7< g\leq1$. (See Section \ref{secRho}.)
So the convexity of $\mathcal{D}_3^{\text{W}}$
restricts Class GHZ to be inside the triangle defined by the vertices
$(g=3/7,w=0)$, $(g=1,w=0)$ and $(g=g_0,w=1-g_0)$.
(Union of tinted domains in Fig.~\ref{figWGHZ}.)
So we can conclude that 
all the states in the blue/grey domain belong to Class GHZ,
and \emph{the border of Class GHZ is in the red/light grey domain of Fig.~\ref{figWGHZ}.}

\section{Conclusions}
\label{secConcl}

In this paper we have investigated the noisy GHZ-W mixture
and demonstrated some necessary but not sufficient criteria
for different classes of separability.
With these criteria we can restrict these classes into some domains of the 2-dimension simplex.
It has turned out that
the strongest conditions was
(\ref{critPPT1}), (\ref{critGS3_11}) and (\ref{critPPT2}) for full separability,
(\ref{critPPT1}) and (\ref{critPPT2}) for the union of Classes 2.8 and 3 and
(\ref{critGS2_1}) and (\ref{critGS2_2}) for biseparability.
These have been obtained from 
the partial transposition criterion of Peres \cite{PeresCrit}
and the criteria of G\"uhne and Seevinck \cite{GuhneSevinckCrit} dealing with matrix elements.
Only these latter criteria have turned out to be strong enough
to reveal a set of entangled states of positive partial transpose.
(The set of these states can be given by the conditions of 
(\ref{critPPT1}), (\ref{critPPT2}) and (\ref{critGS3_11}).
An example is given in Eq.~(\ref{PPTESlo}).)

Besides this, some remarkable coincidences have also appeared:
some parts of some bipartite separability criteria have proved to be necessary and sufficient 
for separability classes of the GHZ-white noise mixture.
(e.g.~the majorisation criterion and the entropy criterion in the $\alpha\to\infty$ limit as well,
and reduction criterion.)
This is interesting because e.g.~the majorisation criterion---as our knowledge---does not state anything about a density matrix
which is majorized by only either of its subsystems.
We do not think that this would be more than a coincidence,
however, the GHZ state is very special
so it is an interesting question 
whether this can be generalized to the $n$-qubit noisy GHZ state.

Another interesting observation was that 
the two settings/measurement vectors strong in detection of GHZ and W state
are related by local unitary Hadamard transformation 
in the criteria on spin-observables of Seevinck and Uffink (Section \ref{secSpin}) 
and also in the criteria of Gabriel \textit{et.~al.}~(Section \ref{secHub}).
The transformation on settings/measurement vectors
can also be written on the state $\varrho\mapsto (H^{\otimes3})^\dagger\varrho H^{\otimes3}$,
which means that we can use the same measurements
on the transformed state
for the detection of W state as for the detection of GHZ state.

We have also investigated the SLOCC classes of fully entangled states
and we have given restrictions for Class GHZ.


\begin{acknowledgments}
We thank P\'eter L\'evay for useful discussions.
Financial support from the T\'AMOP*******
is gratefully acknowledged.
\end{acknowledgments}

\appendix
\section{Matrices}
\label{secMatrices}

In this Appendix we show 
the density matrix $\varrho$ given in Eq.~(\ref{lo}),
its partial transpose,
its reshufflings,
and its marginals explicitly.
Note that we use the renormalized parameters $\td=d/8$, $\tg=g/2$, $\tw=w/3$
and the constraint $d+g+w=1$ holds.
\begin{equation}
\label{MxR}
\varrho=\begin{bmatrix}
 \td+\tg & \cdot   & \cdot   & \cdot   & \cdot   & \cdot   & \cdot   & \tg     \\
 \cdot   & \td+\tw & \tw     & \cdot   & \tw     & \cdot   & \cdot   & \cdot   \\
 \cdot   & \tw     & \td+\tw & \cdot   & \tw     & \cdot   & \cdot   & \cdot   \\
 \cdot   & \cdot   & \cdot   & \td     & \cdot   & \cdot   & \cdot   & \cdot   \\
 \cdot   & \tw     & \tw     & \cdot   & \td+\tw & \cdot   & \cdot   & \cdot   \\
 \cdot   & \cdot   & \cdot   & \cdot   & \cdot   & \td     & \cdot   & \cdot   \\
 \cdot   & \cdot   & \cdot   & \cdot   & \cdot   & \cdot   & \td     & \cdot   \\
 \tg     & \cdot   & \cdot   & \cdot   & \cdot   & \cdot   & \cdot   & \td+\tg 
\end{bmatrix},
\end{equation}
\begin{equation}
\label{MxRT1}
\varrho^{T_1}=\begin{bmatrix}
 \td+\tg & \cdot   & \cdot   & \cdot   & \cdot   & \tw     & \tw     & \cdot   \\
 \cdot   & \td+\tw & \tw     & \cdot   & \cdot   & \cdot   & \cdot   & \cdot   \\
 \cdot   & \tw     & \td+\tw & \cdot   & \cdot   & \cdot   & \cdot   & \cdot   \\
 \cdot   & \cdot   & \cdot   & \td     & \tg     & \cdot   & \cdot   & \cdot   \\
 \cdot   & \cdot   & \cdot   & \tg     & \td+\tw & \cdot   & \cdot   & \cdot   \\
 \tw     & \cdot   & \cdot   & \cdot   & \cdot   & \td     & \cdot   & \cdot   \\
 \tw     & \cdot   & \cdot   & \cdot   & \cdot   & \cdot   & \td     & \cdot   \\
 \cdot   & \cdot   & \cdot   & \cdot   & \cdot   & \cdot   & \cdot   & \td+\tg 
\end{bmatrix},
\end{equation}
\begin{equation}
\label{MxR23}
\varrho^{23}=\begin{bmatrix}
2\td+\tg+\tw & \cdot   & \cdot   & \cdot   \\
 \cdot   &2\td+\tw & \tw     & \cdot   \\
 \cdot   & \tw     &2\td+\tw & \cdot   \\
 \cdot   & \cdot   & \cdot   &2\td+\tg  
\end{bmatrix},
\end{equation}
\begin{equation}
\label{MxR1}
\varrho^1=\begin{bmatrix}
4\td+\tg+2\tw & \cdot   \\
 \cdot   &4\td+\tg+\tw   \\
\end{bmatrix},
\end{equation}
\begin{widetext}
\begin{equation}
\label{MxRRI123}
\Id^1\otimes\varrho^{23}-\varrho=\begin{bmatrix}
 \td+\tw & \cdot   & \cdot   & \cdot   & \cdot   & \cdot   & \cdot   &-\tg     \\
 \cdot   & \td     & \cdot   & \cdot   &-\tw     & \cdot   & \cdot   & \cdot   \\
 \cdot   & \cdot   & \td     & \cdot   &-\tw     & \cdot   & \cdot   & \cdot   \\
 \cdot   & \cdot   & \cdot   & \td+\tg & \cdot   & \cdot   & \cdot   & \cdot   \\
 \cdot   &-\tw     &-\tw     & \cdot   & \td+\tg & \cdot   & \cdot   & \cdot   \\
 \cdot   & \cdot   & \cdot   & \cdot   & \cdot   & \td+\tw & \tw     & \cdot   \\
 \cdot   & \cdot   & \cdot   & \cdot   & \cdot   & \tw     & \td+\tw & \cdot   \\
-\tg     & \cdot   & \cdot   & \cdot   & \cdot   & \cdot   & \cdot   & \td 
\end{bmatrix},
\end{equation}
\begin{equation}
\label{MxRR1I23}
\varrho^1\otimes\Id^{23}-\varrho=\begin{bmatrix}
3\td+2\tw    & \cdot       & \cdot       & \cdot       & \cdot       & \cdot       & \cdot       &-\tg     \\
 \cdot       &3\td+\tg+\tw &-\tw         & \cdot       &-\tw         & \cdot       & \cdot       & \cdot   \\
 \cdot       &-\tw         &3\td+\tg+\tw & \cdot       &-\tw         & \cdot       & \cdot       & \cdot   \\
 \cdot       & \cdot       & \cdot       &3\td+\tg+2\tw& \cdot       & \cdot       & \cdot       & \cdot   \\
 \cdot       &-\tw         &-\tw         & \cdot       &3\td+\tg     & \cdot       & \cdot       & \cdot   \\
 \cdot       & \cdot       & \cdot       & \cdot       & \cdot       &3\td+\tg+\tw & \cdot       & \cdot   \\
 \cdot       & \cdot       & \cdot       & \cdot       & \cdot       & \cdot       &3\td+\tg+\tw & \cdot   \\
-\tg         & \cdot       & \cdot       & \cdot       & \cdot       & \cdot       & \cdot       &3\td+\tw
\end{bmatrix},
\end{equation}
\setcounter{MaxMatrixCols}{16}
\begin{equation}
\label{MxRR24}
R(\varrho)=\begin{bmatrix}
 \td+\tg & \cdot   & \cdot   & \cdot   & \cdot   & \td+\tw & \tw     & \cdot   & \cdot   & \tw     & \td+\tw & \cdot   & \cdot   & \cdot   & \cdot   & \td     \\
 \cdot   & \cdot   & \cdot   & \tg     & \tw     & \cdot   & \cdot   & \cdot   & \tw     & \cdot   & \cdot   & \cdot   & \cdot   & \cdot   & \cdot   & \cdot   \\
 \cdot   & \tw     & \tw     & \cdot   & \cdot   & \cdot   & \cdot   & \cdot   & \cdot   & \cdot   & \cdot   & \cdot   & \tg     & \cdot   & \cdot   & \cdot   \\
 \td+\tw & \cdot   & \cdot   & \cdot   & \cdot   & \td     & \cdot   & \cdot   & \cdot   & \cdot   & \td     & \cdot   & \cdot   & \cdot   & \cdot   & \td+\tg 
\end{bmatrix},
\end{equation}
\begin{equation}
\label{MxRR222}
R'(\varrho)=\begin{bmatrix}
 \td+\tg & \cdot   & \cdot   & \td+\tw & \cdot   & \cdot   & \tw     & \cdot   \\
 \cdot   & \cdot   & \tw     & \cdot   & \cdot   & \tg     & \cdot   & \cdot   \\
 \cdot   & \tw     & \cdot   & \cdot   & \tw     & \cdot   & \cdot   & \cdot   \\
 \td+\tw & \cdot   & \cdot   & \td     & \cdot   & \cdot   & \cdot   & \cdot   \\
 \cdot   & \tw     & \cdot   & \cdot   & \td+\tw & \cdot   & \cdot   & \td     \\
 \tw     & \cdot   & \cdot   & \cdot   & \cdot   & \cdot   & \cdot   & \cdot   \\
 \cdot   & \cdot   & \tg     & \cdot   & \cdot   & \cdot   & \cdot   & \cdot   \\
 \cdot   & \cdot   & \cdot   & \cdot   & \td     & \cdot   & \cdot   & \td+\tg 
\end{bmatrix}.\\
\end{equation}
\end{widetext}

\section{Permutation-invariant states of Classes 2.1 and 2.8}
\label{secExampl}

To see that Class 2.1 is not empty for \emph{permutation-invariant states in general}
we will show an explicit example.
Let $\cket{\beta_0}=(\cket{00}+\cket{11})/\sqrt 2$ be the Bell-state,
then the uniform mixture of the rank one projectors to the subspaces
$\cket{0}_1\otimes\cket{\beta_0}_{23}$, 
$\cket{0}_2\otimes\cket{\beta_0}_{31}$ and
$\cket{0}_3\otimes\cket{\beta_0}_{12}$
gives a state
which is by construction a permutation-invariant $2$-separable one:
\begin{equation}
\varrho^{\text{Class 2.1}}=\frac16\begin{bmatrix}
 3   &\cdot&\cdot& 1   &\cdot& 1   & 1   &\cdot\\
\cdot&\cdot&\cdot&\cdot&\cdot&\cdot&\cdot&\cdot\\
\cdot&\cdot&\cdot&\cdot&\cdot&\cdot&\cdot&\cdot\\
 1   &\cdot&\cdot& 1   &\cdot&\cdot&\cdot&\cdot\\
\cdot&\cdot&\cdot&\cdot&\cdot&\cdot&\cdot&\cdot\\
 1   &\cdot&\cdot&\cdot&\cdot& 1   &\cdot&\cdot\\
 1   &\cdot&\cdot&\cdot&\cdot&\cdot& 1   &\cdot\\
\cdot&\cdot&\cdot&\cdot&\cdot&\cdot&\cdot&\cdot
\end{bmatrix}.
\end{equation}
It can be easily checked that its partial transpose is not positive,
so it is not $\alpha_2$-separable, (the partial transposition criterion see in Section \ref{secPPT})
hence it is in Class 2.1.

An example for a permutation-invariant state in Class 2.8
is given in Eq.~(14) of \cite{Mix3QbSLOCC} with $a=b=\frac1c$:
\begin{equation} 
\varrho^{\text{Class 2.8}}=\frac1{2+3\left(a+\frac1a\right)}\begin{bmatrix}
 1   &\cdot&\cdot&\cdot&\cdot&\cdot&\cdot& 1   \\
\cdot& a   &\cdot&\cdot&\cdot&\cdot&\cdot&\cdot\\
\cdot&\cdot& a   &\cdot&\cdot&\cdot&\cdot&\cdot\\
\cdot&\cdot&\cdot&\frac1a&\cdot&\cdot&\cdot&\cdot\\
\cdot&\cdot&\cdot&\cdot& a   &\cdot&\cdot&\cdot\\
\cdot&\cdot&\cdot&\cdot&\cdot&\frac1a&\cdot&\cdot\\
\cdot&\cdot&\cdot&\cdot&\cdot&\cdot&\frac1a&\cdot\\
 1   &\cdot&\cdot&\cdot&\cdot&\cdot&\cdot& 1   \\
\end{bmatrix},
\end{equation}
where $0<a$.
This state is entangled if and only if $a\neq1$,
and $\varrho^{\text{Class 2.8}}\in \mathcal{D}_3^{\alpha_2}$ for all $\alpha_2$.~\cite{Mix3QbSLOCC}

\section{Wootters concurrence}
\label{secWoott}

For mixed states the only measure of entanglement known explicitly is the 
\emph{Wootters-concurrence} of two qubits: 
\cite{HillWoottersConc,WoottersConc}
\begin{equation}
\label{conc}
\mathcal{C}(\omega)=\max\{0,\lambda^\downarrow_1-\lambda^\downarrow_2-\lambda^\downarrow_3-\lambda^\downarrow_4\},
\end{equation}
where $\omega$ is a two-qubit density matrix 
and $\lambda^\downarrow_i$s are the eigenvalues of the positive matrix 
$\sqrt{\sqrt{\omega}\tilde{\omega}\sqrt{\omega}}$ in non-increasing order.
Here $\tilde{\omega}$ denotes the \emph{spin-flipped} density matrix: 
$\tilde{\omega}=(\sigma_2\otimes\sigma_2)\omega^*(\sigma_2\otimes\sigma_2)$
where $\sigma_2=\left[\begin{smallmatrix}0&-i\\i&0\end{smallmatrix}\right]$ and $\omega^*$ denotes the complex conjugation.
Equivalently $\lambda_i$s can be calculated as the square roots of the positive eigenvalues of 
the matrix $\tilde{\omega}\omega$.

\begin{figure}[h]
 \centering
 \includegraphics[width=0.8\columnwidth]{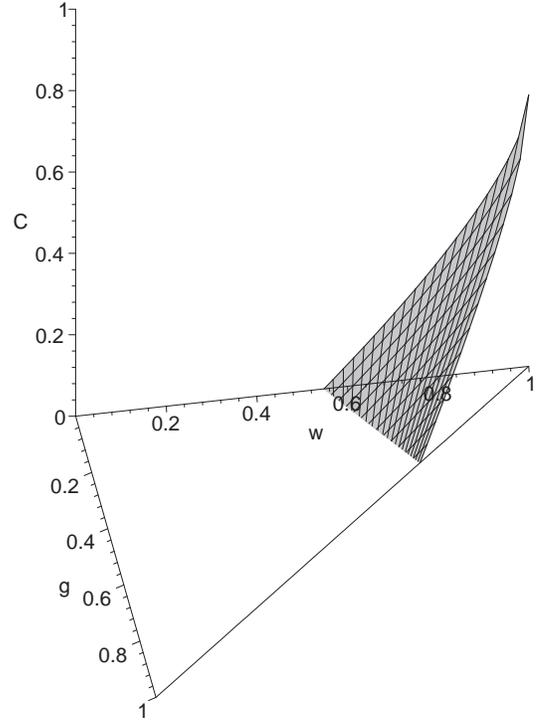}
 \caption{Wootters-concurrence of $\varrho^{23}$ on the $g$-$w$-plane. (Eq.~(\ref{Conc}))}
 \label{figWoott}
\end{figure}

Let us calculate now the concurrence given in Eq.~(\ref{conc}) of the matrix $\varrho^{23}$.
(See in Eq.~(\ref{MxR23}) of the Appendix.)
Since the spin-flip in the two-qubit case means 
transpose with respect to the antidiagonal
then multiplication of neither diagonal nor antidiagonal entries by $-1$,
one can easily get:
\begin{multline}
\label{spectRho23W}
\Spect(\tilde{\varrho}^{23}\varrho^{23})=\\
\begin{aligned}[t]\bigl\{\;
4(\td+\tw)^2&= 4(3-3g+5w)^2/24^2,\\
(2\td+\tg)&(2\td+\tg+\tw)\\
&= 12(1+g-w)(3+3g+w)/24^2,\\
(2\td+\tg)&(2\td+\tg+\tw)\\
&= 12(1+g-w)(3+3g+w)/24^2,\\
4\td^2&= 36(1-g-w)^2/24^2\;\bigr\}
\end{aligned}
\end{multline}
Clearly, the last eigenvalue is the smallest one.
If $4(\td+\tw)^2\leq(2\td+\tg)(2\td+\tg+\tw)$ 
then $\lambda^\downarrow_1=\lambda^\downarrow_2$ hence $\mathcal{C}(\varrho^{23})=0$.
If $4(\td+\tw)^2\geq(2\td+\tg)(2\td+\tg+\tw)$, 
then $\lambda^\downarrow_2=\lambda^\downarrow_3$ and $\mathcal{C}(\varrho^{23})$ can be nonzero.
It turns out that
\begin{equation}
\label{Conc}
\begin{split}
\mathcal{C}(\varrho^{23})&=2\tw-2\sqrt{(2\td+\tg)(2\td+\tg+\tw)}   \\
&=\frac{2}{3}w-\frac{1}{2\sqrt{3}}\sqrt{(1+g-w)(3+3g+w)}
\end{split}
\end{equation}
if $0\leq \tw^2 - (2\td+\tg)(2\td+\tg+\tw)
=(-9g^2+19w^2+6gw-18g+6w-9)/12^2$,
otherwise $\mathcal{C}(\varrho^{23})=0$. (Fig.~\ref{figWoott})
It takes its maximum $2/3$ in $g=0$, $w=1$, i.e.~in pure W-state.
For the GHZ-W mixture ($d=0$) we get back the result of Ref.~\cite{MixedThreetangle}.

\bibliography{criteria4}

\end{document}